\documentclass[aps,prc,preprint,groupedaddress,showpacs]{revtex4}

\usepackage{graphicx}
\usepackage{epstopdf}
\usepackage{amsmath}

\newcommand {\be}{\begin{equation}}
\newcommand {\ee}{\end{equation}}
\newcommand {\bea}{\begin{eqnarray}}
\newcommand {\eea}{\end{eqnarray}}

\newcommand{\btau}{\mbox{\boldmath$\tau$}}
\newcommand{\bpi}{\mbox{\boldmath$\pi$}}

\newcommand{\lqt}{\textquotedblleft}
\newcommand{\la}{\langle}
\newcommand{\ra}{\rangle}
\newcommand{\cgz}{\left\la1/2\ m_1,1/2\ m_2\mid0\ 0\right\ra}

\begin{document}

\preprint{NT@UW-09-13}

\title{Charge Symmetry Breaking in the $np\rightarrow d\pi^0$ reaction}

\author{Daniel R. Bolton}
\author{Gerald A. Miller}
\affiliation{Department of Physics, University of Washington, Seattle, Washington 98195-1560, USA}

\date{\today}

\begin{abstract}
The asymmetry in the angular distribution of $np\rightarrow d\pi^0$ due to Charge Symmetry Breaking is calculated using Heavy Baryon Chiral Perturbation Theory.  Recent developments in power counting have proven successful in describing total cross sections, and we apply them to the asymmetry calculation.  Reducibility in one of the leading order diagrams is examined.  We compare the updated theory with experimental results for a range of physically reasonable parameters and find over-prediction for the entire range.
\end{abstract}

\pacs{11.30.Er, 24.80.+y, 25.10+s}

\maketitle

\section{Introduction\label{sec:introduction}}

The nuclear reaction $NN\rightarrow NN\pi$ has been studied for a long time in many ways.  Theoretical understanding of the reaction is currently pursued using Heavy Baryon Chiral Perturbation Theory (HB$\chi$PT) \cite{Jenkins:1990jv, Bernard:1993nj, Hemmert:1997ye, Bedaque:2002mn}.  This effective theory of the strong nuclear force treats hadrons and mesons as the fundamental degrees of freedom.  The Lagrangian contains an infinite number of terms which decrease in importance according to an expansion in a (somewhat) small parameter.  Such an understanding is important because it is model independent, provided convergence is achieved.  Furthermore, once theorists arrive at an ordering scheme for the expansion, error estimates will become more reliable and calculations of different and more exotic reactions possible.

One particularly interesting observable that HB$\chi$PT can help determine is the magnitude of Charge Symmetry Breaking (CSB) \cite{Cheung:1979ma, Epelbaum:1999zn, Miller:2006tv}.  Charge symmetry refers to the approximate invariance of hadronic systems under an isospin rotation of $\pi$ about the y-axis.  This symmetry is broken by the mass difference between up and down quarks and by electromagnetic effects \cite{Slaus:1990nn}.  We will discuss the corresponding symmetry breaking terms in the HB$\chi$PT Lagrangian.  These terms are especially important for the interactions of neutral pions \cite{Weinberg:1977}.

Relations involving the light quark masses can be obtained from the SU(3) chiral symmetry on which HB$\chi$PT is based, but experiments which directly measure $m_d-m_u$ are difficult because of the lack of a neutral pion beam.  One experiment which overcomes this difficulty and also minimizes electromagnetic effects is $np\rightarrow d\pi^0$.  The angular distribution of this reaction is symmetric about $90^\circ$ in the center of mass when charge symmetry is respected.  Ref. \cite{Opper:2003sb} recently observed that this distribution is asymmetric.

This report advances previous work in several ways.  The authors of Ref. \cite{Lensky:2005jc} showed that a vertex which was thought to be higher order in fact contributes at leading order.  This lead to a much improved understanding of the total cross section of $pp\rightarrow d\pi^+$.  We extend this calculation off threshold for neutral pion production and make a comparison with the corresponding data.  We also investigate a subtlety regarding reducibility in one of the leading order diagrams and comment on the resulting corrections.  Finally, we calculate the asymmetry of $np\rightarrow d\pi^0$, bringing up to date the calculations of Ref. \cite{Niskanen:1998yi} and \cite{vanKolck:2000ip} by including the new effects discussed in this paragraph.

In Sec. \ref{sec:kinematics} we discuss the kinematics and selection rules of the $np\rightarrow d\pi^0$ process.  Then, in Sec. \ref{sec:cross section} we develop the formalism necessary for calculation of the cross section.  Here we detail the procedure for handling strongly interacting initial and final states.  In Sec. \ref{sec:strong} we present the leading order diagrams with vertices from the isospin conserving part of the Lagrangian (Appendix \ref{sec:lagrangian}).  In this section we also provide a review of the power counting developed by Ref. \cite{Lensky:2005jc} and discuss its impact on neutral pion production.  Next, in Sec. \ref{sec:csb}, we present diagrams with vertices from the CSB part of the Lagrangian which contribute to the reaction.  Our results are given in Sec. \ref{sec:results} and discussed in Sec. \ref{sec:discussion}.  Also in Sec. \ref{sec:discussion}, we compare our work with another recent calculation of the asymmetry by Ref. \cite{Filin:2009yh}.

\section{Kinematics and Selection Rules\label{sec:kinematics}}

At threshold, the reaction $NN\rightarrow NN\pi$ produces a pion from the kinetic energy of the incoming NN pair.  Let $\vec{p}=\frac{1}{2}(\vec{p}_1-\vec{p}_2)$ be the relative momentum of the pair.  In the center of mass frame for this reaction, the total momentum $\vec{P}=\vec{p}_1+\vec{p}_2=0$ and thus $\vec{p}_1=\vec{p}$ and $\vec{p}_2=-\vec{p}$.  In the non-relativistic limit, the total initial energy is $E_i=2M_N+p^2/M_N$ where $M_N$ is the nucleon mass (the average mass is used for $np$ reactions).  The energy of the final state at threshold is $E_f=2M_N-E_b+m_\pi$ where $E_b=2.224$ MeV is the deuteron binding energy and $m_\pi$ is the mass of the appropriate pion.  Neglecting the binding energy yields a simple scale for external momenta in these reactions, $\tilde{p}\equiv\sqrt{m_\pi M_N}=356$ MeV.

There are two frames to keep in mind: the center of mass frame (\textbf{C}), and the lab frame where the proton is at rest (\textbf{E} for experiment).  The experiment of interest \cite{Opper:2003sb} is $np\rightarrow d\pi^0$ performed in the \textbf{E} frame.  In this frame the invariant is expressed as $s_\mathbf{E}=(M_p+M_n)^2+2M_pT_L$ where $T_L$ is the kinetic energy defined by $E_n=M_n+T_L$.  The experiment was performed at $T_L=279.5$ MeV, slightly above the threshold value of $T_L=275.1$ MeV.  To simplify the formalism, we use the \textbf{C} frame to do the calculation.  In terms of the pion momentum, $\vec{q}$, and the deuteron mass, $M_d$, we have $\sqrt{s_\mathbf{C}}=\sqrt{M_d^2+\vec{q}\,^2}+\sqrt{m_\pi^2+\vec{q}\,^2}$.  It is convenient to define a dimensionless parameter $\eta$ to describe how far off threshold the reaction is, $\eta=|\vec{q}\,|/m_\pi$.  Equating the invariants we find $\eta_\mathbf{C}=0.169$ ($q_\mathbf{C}=22.86$ MeV) at the experimental energy. 

We use the hybrid approach developed by Weinberg \cite{Weinberg:1991um} to calculate the cross section, due to the fact that the initial and final states are strongly interacting.  The \lqt operator" is calculated as the sum of two-particle-irreducible diagrams involving four nucleon lines (two incoming and two outgoing) and one pion line.  Then, a phenomenological potential is used to calculate the NN scattering wavefuctions and the deuteron bound state wavefunction.  Finally, the operator is convolved with the wavefunctions to obtain the matrix element.  This method is described in detail in Sec. \ref{sec:cross section}, but first we present the selection rules for $np\rightarrow d\pi^0$.

The deuteron has the following quantum numbers: spin $S=1$, orbital angular momentum $L=0,2$ (parity even), total angular momentum $J=1$, and isospin $T=0$.  If the pion ($J^\pi=0^-$) is produced in the s-wave, then conservation of total angular momentum gives $J_i=1$.  Furthermore, since the pion is parity odd, s-wave pions must be produced from a parity odd $np$ wavefunction, $L_i=1$.  Likewise, if the pion comes in the p-wave, the initial parity is even and $J_i=0,1,2$.

To completely pin down the quantum numbers of the initial state, we turn to isospin.  If isospin is conserved in the final state, then $T_f=1$ because the deuteron is an isoscalar.  \lqt Strong" diagrams have $T_i=1$ with the isospin part of the initial wavefunction symmetric under exchange of nucleons.  Overall wavefunction antisymmetry then requires $S_i=1$ for $l_\pi=0$ and $S_i=0$ for $l_\pi=1$.  Thus the initial $np$ pair must be in the $^3P_1$ channel for $l_\pi=0$, while the available channels become $^1S_0$ and $^1D_2$ for $l_\pi=1$.  CSB operators transform as vectors under isospin, thus the initial neutron-proton state must have $T_i=0$.  The same arguments as for the strong operators then show that s-wave pions are produced from $^1P_1$ $np$ pairs while p-wave pions are produced from the coupled channels $^3S_1$ and $^3D_1$ in addition to $^3D_2$.  These conclusions are summarized in Table \ref{tab:channels}.
\begin{table}
\caption{\label{tab:channels}Channels for the $np$ wavefunction in $np\rightarrow d\pi^0$}
\begin{center}
\begin{tabular}{|c|c|c|}
\hline
& Strong & CSB \\ \hline
$l_\pi=0$ & $^3P_1$ & $^1P_1$ \\ \hline
$l_\pi=1$ & $^1S_0$, $^1D_2$ & $^3S_1$, $^3D_1$, $^3D_2$ \\ \hline
\end{tabular}
\end{center}
\end{table}

The observable of interest in the experiment is the forward/backward asymmetry in the differential cross-section given by
\be
A_{fb}=\frac{\int_0^{\pi/2}d\Omega\ [\sigma(\theta)-\sigma(\pi-\theta)]}{\int_0^\pi d\Omega\ \sigma(\theta)}.
\label{eq:asymmetry}
\ee
A non-zero asymmetry will only be observed when initial states of opposite parity interfere.  However, the interference can only occur for states with the same spin since the spin z-components get summed over.  Thus for calculating the asymmetry, we are concerned with two terms: (s-wave strong)$\cdot$(p-wave CSB) and (p-wave strong)$\cdot$(s-wave CSB).

\section{Cross Section\label{sec:cross section}}

\subsection{Cross Section Formalism\label{sec:formalism}}

We will now derive an expression for the cross section in the \textbf{C} frame where the pion momentum is $\vec{q}$.  First we must define expressions for the strongly interacting initial and final states.  To form an interacting NN state with total momentum $\vec{P}=\vec{p_1}+\vec{p_2}$, we use a superposition of free particle states
\bea
\left |\psi(\vec{P})\right\ra&=&\int\frac{d^3p_1}{(2\pi)^3}\frac{d^3p_2}{(2\pi)^3}\ \psi\left(\frac{|\vec{p}_1-\vec{p}_2|}{2}\right)\ \left |N(\vec{p}_1),N(\vec{p}_2)\right\ra\delta(\vec{P}-\vec{p}_1-\vec{p}_2)\nonumber
\\
&=&\int\frac{d^3p}{(2\pi)^3}\ \psi(p)\ \left |N(\vec{p}+\vec{P}/2),N(-\vec{p}+\vec{P}/2)\right\ra,
\eea
where spin and isospin have been ignored for now.  The wavefunction $\psi(p)$ is obtained by solving the Schr\"{o}dinger equation with the appropriate NN potential specified below.  In the \textbf{C} frame, $\vec{P}_i=0$ and if the nucleons forming the deuteron have momentum $\vec{k}_{1,2}$, then $\vec{k}_1+\vec{k}_2\equiv\vec{K}=-\vec{q}$.  The invariant matrix element is then
\be
\mathcal{M}\left(N(p_1),N(p_2))\rightarrow \pi(q),d(K)\right)=\int\frac{d^3k}{(2\pi)^3}\frac{d^3p}{(2\pi)^3}\ \psi_d^*(k)\ \hat{\mathcal{M}}\left(p,k,q\right)\ \psi_{np}(p),
\label{eq:mxel}
\ee
where $p_{1,2}=(E_{1,2},\pm\vec{p}\,)$, $q=(\omega_q,\vec{q}\,)$, $K=(E_d,-\vec{q}\,)$.  Note that we are treating the initial state as two separate particles, but the deuteron as a single particle.  As mentioned in Sec. \ref{sec:kinematics}, the sum of the two-particle-irreducible diagrams with external momenta $\vec{q},\vec{p}_{1,2},\vec{k}_{1,2}$ (the \lqt operator") is denoted $\hat{\mathcal{M}}$, and is convolved with the external wavefunctions.  The operator is calculated in momentum space and is a function of the external momenta $p,k,$ and $q$.  Also note that the wavefunctions will include spin and isospin kets on which the operator acts.

We perform the calculation in position space by inserting Fourier representations of both wavefunctions
\be
\psi_{NN}(\vec{r})=\int\frac{d^3p}{(2\pi)^3}\ e^{i\vec{p}\cdot\vec{r}}\psi_{NN}(\vec{p}).
\label{eq:psift}
\ee
A Fourier representation of the operator with respect to $\vec{l}\equiv\vec{k}-\vec{p}$ is also inserted
\be
\hat{\mathcal{M}}\left(\vec{r}\right)=\int\frac{d^3l}{(2\pi)^3}\ e^{i\vec{l}\cdot\vec{r}}\hat{\mathcal{M}}\left(\vec{l},\vec{q}\right).
\label{eq:mft}
\ee
As described in Appendix \ref{sec:method}, $\vec{l}$ is the momentum that appears in pion production reactions: $\mathcal{M}(\vec{p},\vec{k},\vec{q})\rightarrow\mathcal{M}(\vec{l},\vec{q})$.

Now we can rewrite Eq. (\ref{eq:mxel})
\bea
\mathcal{M}&=&\int d^3r\ d^3r'\ d^3r''\ \psi_d^*(r'')\ \hat{\mathcal{M}}\left(\vec{r}\right)\ \psi_{np}(r')\ \int\frac{d^3k}{(2\pi)^3}\frac{d^3p}{(2\pi)^3}\ e^{i\vec{k}\cdot\vec{r}\,''}e^{-i(\vec{k}-\vec{p})\cdot\vec{r}}e^{-i\vec{p}\cdot\vec{r}\,'}\nonumber
\\
&=&\int d^3r\ \psi_d^*(r)\ \hat{\mathcal{M}}\left(\vec{r}\right)\ \psi_{np}(r).
\label{eq:m}
\eea
With these choices of Fourier representations, the momentum integrals evaluate to delta functions which allow evaluation of the spatial integrals.  The result is an integral over a single spatial variable.

Next, we express the invariant S-matrix element in terms of $\mathcal{M}$
\be
\left\la\pi^0(q)d(K)\mid S\mid p(p_1),n(p_2)\right\ra=1-i(2\pi)^4\delta^4(q+K-p_1-p_2)\mathcal{M}(p_1,p_2\rightarrow q,K).
\label{eq:smatrix}
\ee
In the center of mass frame, the differential cross section is
\be
d\sigma=\frac{1}{4|\vec{p}\,|\sqrt{s}}\frac{1}{4}\sum_{m_d,m_1,m_2}\left|\mathcal{M}\right|^2(2\pi)^4\delta^4(q+K-p_1-p_2)\frac{d^3q}{(2\pi)^32\omega_q}\frac{d^3K}{(2\pi)^32E_d},
\ee
where we have averaged over the four $np$ spin states and summed over the three spin states of the deuteron.  The vector part of the delta function tells us $\vec{q}=-\vec{K}$ and the energy part tells us $E_1+E_2=\omega_q+E_d$.  This allows us to perform all but the $d\Omega_K$ integral,
\be
\frac{d\sigma}{d\Omega}=\frac{|\vec{q}\,|}{64\,\pi^2\,s\,|\vec{p}\,|}\frac{1}{4}\sum_{m_d,m_1,m_2}|\mathcal{M}|^2.
\label{eq:dsdoab}
\ee
What remains is to derive expressions for the wavefunctions and the operator appearing in Eq. (\ref{eq:m}).

\subsection{Initial and Final States\label{sec:states}}

In the absence of interactions the wavefunction $\psi_{np}(\vec{r}\,)$ is determined by performing a partial wave expansion on an antisymmetrized wavefunction of a free proton and a free neutron with relative momentum $\vec{p}$.  First we consider the strong operators where the $np$ pair is in an isospin-1 state,
\bea
\left(\vec{r}\mid\psi_{np}\right\ra&=&\mathcal{P}_{T=1}\frac{1}{\sqrt{2}}\left(e^{i\vec{p}\cdot\vec{r}}|m_1,m_2\ra\otimes|T_{z,1},T_{z,2}\ra-e^{-i\vec{p}\cdot\vec{r}}|m_2,m_1\ra\otimes|T_{z,2},T_{z,1}\ra\right)\nonumber
\\
&=&\frac{1}{\sqrt{2}}\left(e^{i\vec{p}\cdot\vec{r}}|m_1,m_2\ra-e^{-i\vec{p}\cdot\vec{r}}|m_2,m_1\ra\right)\otimes\frac{1}{\sqrt{2}}\left|T=1,T_z=0\right\ra,
\label{eq:npwfn}
\eea
where $\mathcal{P}_{T=1}$ is the isospin projector.  The bra $\left(\vec{r}\;\right|$ indicates that we are choosing a basis for space, but not for spin or isospin.  Implicit in the notation is the dependence of $(\vec{r}\mid\psi_{np}\ra$ on the momentum $\vec{p}$, the spin z-components of the two nucleons, $m_i$, and the isospin z-components of the two nucleons, $T_{z,i}$, with the requirement that $T_{z,1}+T_{z,2}=0$.

The exponentials are now expanded and the presence of the strong interaction is added by changing the spherical Bessel functions, $j_L(pr)\rightarrow e^{i\delta_{L}}u_{L,J}(r)/pr$.  The $u_{L,J}$ functions and the $\delta_{L}$ phase shifts are obtained by solving the Schr\"odinger equation with a phenomenological NN potential (we use Argonne V18 \cite{Wiringa:1994wb}).  Finally, the spherical harmonics are combined with the spin kets to form states with definite total angular momentum.  The notation for these states is $\left|(SL)J,m_J\right\ra\otimes\mid T,T_z\ra$.  For the allowed quantum numbers, we find
\bea
\lefteqn{\left(\vec{r}\mid\psi_{np}(^{2S+1}L_J,T=1)\right\ra=4\pi (i)^Le^{i\delta_{L}}\frac{u_{L,J}(r)}{pr}\left\la1/2\ m_1,1/2\ m_2\mid S\ m_s\right\ra\nonumber}
\\[0.1in]
&\times&\sum_{m_i}\left\la S\ m_s,L\ m_i-m_s\mid J\ m_i\right\ra Y^{L\ *}_{m_i-m_s}(\hat{p})\left(\hat{r}\mid(SL)J,m_i\right\ra\otimes\left|1,0\right\ra,
\label{eq:npwfnT1}
\eea
where $m_s=m_1+m_2$ and the second Clebsch Gordan coefficient allows us to make the sum over $m_i=m_l+m_s$ rather than $m_l$.  For the CSB operators, we have $T=0$ np wavefunctions and find
\bea
\lefteqn{\left(\vec{r}\mid\psi_{np}(^{2S+1}L_J,T=0)\right\ra=\pm4\pi (i)^Le^{i\delta_{L}}\frac{u_{L,J}(r)}{pr}\left\la1/2\ m_1,1/2\ m_2\mid S\ m_s\right\ra\nonumber}
\\[0.1in]
&\times&\sum_{m_i}\left\la J\ m_i\mid S\ m_s,L\ m_i-m_s\right\ra Y^{L\ *}_{m_i-m_s}(\hat{p})\left(\hat{r}\mid(SL)J,m_i\right\ra\otimes\left|0,0\right\ra,
\label{eq:npwfnT0}
\eea
where the $\pm$ refers to $T_{z,1}=\pm1/2$.  Similar analysis gives the final-state wavefunction of the deuteron as a function of its polarization, $m_f$,
\be
\left\la\psi_d(m_f)\mid\vec{r}\,\right)=\left\la0,0\right|\otimes\left(\frac{u(r)}{r}\left\la(10)1,m_f\mid\hat{r}\right)+\frac{w(r)}{r}\left\la(12)1,m_f\mid\hat{r}\right)\right).
\label{eq:dwfn}
\ee

\section{Strong Contribution\label{sec:strong}}

\subsection{Diagrammatic Expansion\label{sec:diagrams}}

Before we calculate the effects of charge symmetry, we need to discuss the power counting scheme which leads to a calculation of the total cross section in agreement with experiment.  HB$\chi$PT orders contributions in powers of the external momenta divided by the chiral symmetry breaking scale, which is $\sim M_N$ \cite{Lensky:2005jc}.  In this inelastic reaction, both $q$ and $\tilde{p}$ appear as external momenta and we need to keep track of both in the power counting.  We define the expansion parameter $\chi\equiv\tilde{p}/M_N=\sqrt{m_\pi/M_N}=0.40$.

The Lagrangian is given in Appendix \ref{sec:lagrangian}.  The index of a \lqt type $i$" vertex is given by
\be
\nu_i=d_i+\frac{f_i}{2}-2,
\ee
where $d_i$ is the sum of the number of derivatives, $m_\pi$'s, and $\delta$'s (the $\Delta$N mass difference), and $f_i$ is the number of fermion fields.  In standard power counting at tree level, the sum of the $\nu_i$ for each vertex in a diagram indicates the power of $\chi$ at which that diagram contributes.  This rule, however, will require modification due to the relatively large value of $p$.

\begin{figure}
\centering
\includegraphics[height=1in]{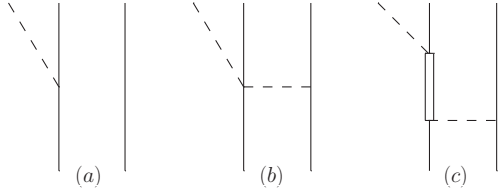}
\caption{Leading order contributions to $np\rightarrow d\pi^0$.  Solid lines represent nucleons, the double solid line represents a $\Delta$, and dashed lines represent pions.\label{fig:stronglo}}
\end{figure}

There are three two-particle irreducible diagrams which can be drawn using the vertices from $\mathcal{L}^{(0)}$.  They will be referred to as the impulse (Fig. \ref{fig:stronglo}a), rescattering (Fig. \ref{fig:stronglo}b), and Delta (Fig. \ref{fig:stronglo}c) diagrams.
There is some ambiguity as to what is meant by Fig. \ref{fig:stronglo}a.  We will discuss this in detail below.  First let us make a few observations about these three diagrams.  We have stated that the initial relative momentum is large.  The final state nucleons (the deuteron) have a comparably small relative momentum.  This \lqt momentum mismatch" is provided for in the rescattering diagram, but not in the impulse diagram.  For this reason, the rescattering diagram dominates the total cross section.  This diagram is strongest in the channel with s-wave pions, and the $\pi\pi NN$ vertex is typically referred to as the Weinberg-Tomozawa (WT) vertex.  Next, although the Delta diagram provides the momentum transfer required, the Delta resonance is at $1232$ MeV and the $\pi N$ energy is $\approx 1080$ MeV, so the Delta diagram is also somewhat suppressed for our situation of interest.  Finally, we note that both the impulse and the Delta diagrams are strongest for the channels in which the pion is in a p-wave.

\begin{figure}
\centering
\includegraphics[height=1in]{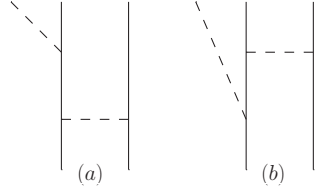}
\caption{Kinematically correct, but reducible impulse contributions.  Solid lines represent nucleons and dashed lines represent pions.\label{fig:strongia}}
\end{figure}

\begin{figure}
\centering
\includegraphics[height=1in]{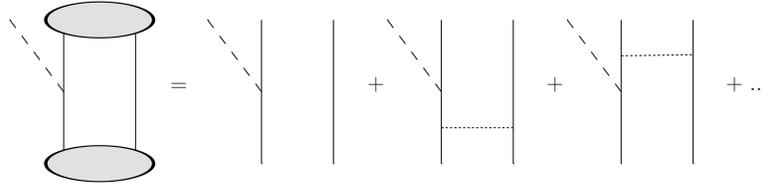}
\caption{\lqt Hybrid" approach.  Solid lines represent nucleons, dashed lines represent pions, dotted lines represent pions with $\omega=0$, and filled ovals represent NN strong interactions.\label{fig:strongiawfn}}
\end{figure}

We now return to the issue of the impulse diagram.  On the one hand, we know that a single nucleon cannot emit a pion and remain on shell.  But on the other hand, the diagrams in Fig. \ref{fig:strongia} which remedy this problem by including one pion exchange (OPE) appear to be two particle reducible.
To resolve this issue we first note that near threshold the energy of the exchanged pion in each of these diagrams is approximately $\omega\approx m_\pi/2$.  However, the diagram of Fig. \ref{fig:stronglo}a is evaluated as diagrams of the form of Fig. \ref{fig:strongia}.  This is due to hybrid nature of the calculation; once the operator (traditionally, the irreducible diagram) is calculated, it is convolved with NN wavefunctions.  One of the major terms of the strong interaction potential at low energy arises from static OPE ($\omega=0$).  We ignore the effects of the rest of the wavefunction for the moment.  The effects of static OPE are schematically shown in Fig. \ref{fig:strongiawfn}.

\begin{figure}
\centering
\includegraphics[width=6in]{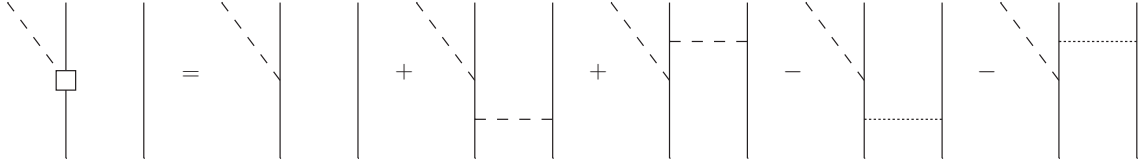}
\caption{Complete impulse contribution.  Solid lines represent nucleons, dashed lines represent pions, dotted lines represent \lqt wavefunction" pions with $\omega=0$, and the square represents the operator which is used for the full impulse approximation.\label{fig:strongiasum}}
\end{figure}

To obtain the correct impulse contribution, we add up the contributions from Figs. \ref{fig:stronglo}a and \ref{fig:strongia} and then subtract what is already included in the wavefunctions (the last two diagrams of Fig. \ref{fig:strongiawfn}).  This calculation is schematically shown in Fig. \ref{fig:strongiasum}.

The OPE propagator is of Yukawa form,
\be
D_\pi(r)=-\frac{e^{-\mu(\omega) r}}{4\pi r},
\ee
where $\mu(\omega)=\sqrt{m_\pi^2-\omega^2}\approx\sqrt{3}/2m_\pi$.  Thus subtracting off the final two diagrams in Fig. \ref{fig:strongiasum} amounts to making the replacement
\be
\frac{e^{-\sqrt{3}m_\pi r/2}}{r}\rightarrow\frac{e^{-\sqrt{3}m_\pi r/2}}{r}-\frac{e^{-m_\pi r}}{r},
\ee
in the exchanged pion propagator.  The final four terms of Fig. \ref{fig:strongiasum} comprise a correction to the impulse diagram, which we will refer to as a \lqt wavefunction correction."  We find that this correction is $\sim4\%$ of the total impulse amplitude at the experimental energy, and we include it in our calculation.

\subsection{Power Counting\label{sec:power counting}}

Now we will look more closely at the size of these diagrams in the \lqt$\tilde{p}$ kinematics", using the counting techniques developed in Ref. \cite{Hanhart:2003pg}.  The propagators are calculated from the Lagrangian in Eq. (\ref{eq:l0})
\bea
D_N(p)&=&\frac{i}{p^0+i\epsilon}\nonumber
\\
D_\Delta(p)&=&\frac{i}{p^0-\delta+i\epsilon}\nonumber
\\
D_\pi(p)&=&\frac{-i}{\vec{p}\,^2+(m_\pi^2-(p^0)^2)-i\epsilon}.\label{eq:props}
\eea
Both the $\pi NN$ and the $\pi N\Delta$ vertices have a momentum dependence of $|\vec{q}\,|$, the pion momentum \textit{at that vertex}.  Note that this momentum is $\tilde{p}$ in the OPE verticies.  The WT vertex contains a factor of $\omega_{q,in}+\omega_{q,out}$.

The external particles have the same momenta in each diagram.  The produced pion has $q=(\omega_q,\vec{q}\,)\approx(m_\pi,0)$, and the incoming nucleons have $p_{1,2}=(E_{1,2},\pm\vec{p}\,)\approx\left(m_\pi/2,\pm\vec{\tilde{p}}\,\right)$ in the Heavy Baryon formalism in which the nucleon mass is subtracted off of the energy component.  Consider the impulse diagram of Fig. \ref{fig:strongia}a.   The final emission contributes $q$, the nucleon propagator $1/m_\pi$, and the OPE $\tilde{p}\cdot1/\tilde{p}^2\cdot\tilde{p}$, so that the diagram is $\sim q/m_\pi=\eta$.  The rescattering diagram is $\sim\frac{3m_\pi/2}{\tilde{p}}\sim\chi$, and the Delta diagram is $\sim\frac{q}{m_\pi-\delta}$.  Finally, note that $\eta\approx\chi^2$ and $\delta\approx 2m_\pi$ so that while the rescattering diagram is $\sim\chi$ the impulse and Delta diagrams are numerically $\sim\chi^2$.  This ordering comes in agreement with the fact that the rescattering diagram is strongest for s-wave pions while the impulse and Delta diagrams are strongest for p-wave pions.

It is well documented that these three diagrams alone do not correctly reproduce the experimental data for the reaction; see Ref. \cite{Hanhart:1995ut} for a review of the theory of meson production.  Near threshold ($\eta\approx0.05$), the most recent experiment found $\alpha_{exp}(np\rightarrow d\pi^0)\approx90\ \mu$b \cite{Hutcheon:1989bt}, while for these first three diagrams, we find $\alpha\approx55\ \mu$b.  The solution to this problem was discovered by Ref. \cite{Lensky:2005jc}, who noticed that the $\nu=1$ \lqt recoil" correction to the WT vertex, which is found in Eq. (\ref{eq:l1}) goes like $(\vec{q}_{in}+\vec{q}_{out})\cdot(\vec{p}_{in}+\vec{p}_{out})/(2M_N)$ where $\vec{q}$ is the pion momentum and $\vec{p}$ is the nucleon momentum.
\begin{figure}
\centering
\includegraphics[height=1in]{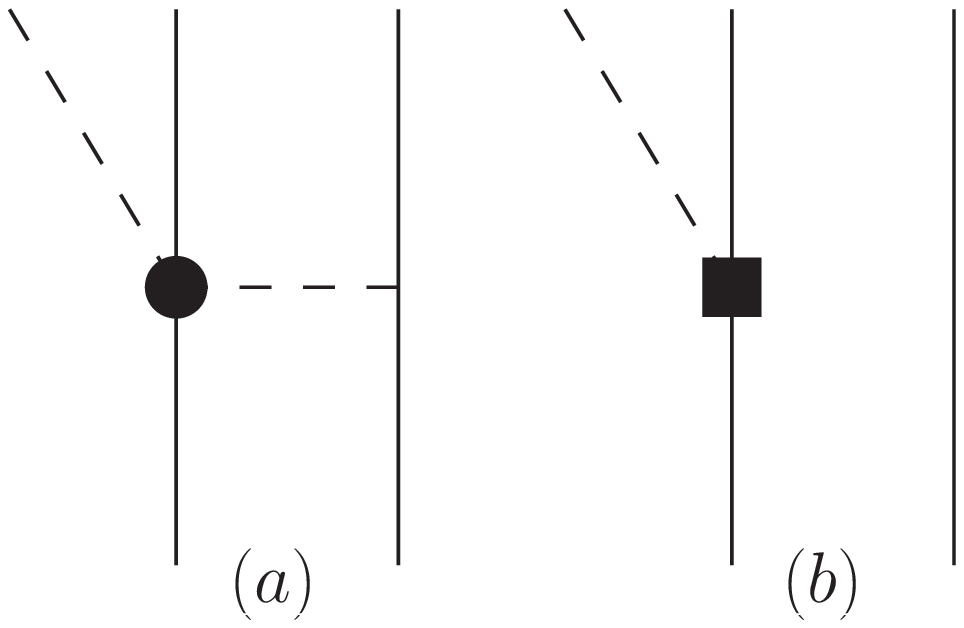}
\caption{Recoil corrections.  Solid lines represent nucleons, dashed lines represent pions, and the filled circle and square represent $\nu=1$ vertices.\label{fig:strongrecoil}}
\end{figure}
For Fig. \ref{fig:strongrecoil}a, this vertex (a filled circle) $\sim\vec{\tilde{p}}^{\ 2}/(2M_N)=m_\pi/2$ and thus this diagram is of order $\chi$, the same order as the $\nu=0$ rescattering diagram.  Similarly, we find that the s-wave portion of the recoil correction to the impulse diagram (Fig. \ref{fig:strongrecoil}b) is of order $\chi$.  In this diagram, the filled square represents the sum analogous to Fig. \ref{fig:strongiasum} for the recoil diagram. We find that the wavefunction corrections are more important ($\sim20\%$) in this case.  Finally, the s-wave portion of the Delta diagram's recoil correction is found to be higher order and is therefore ignored.

The recoil corrections to the propagators have also been included in the calculation where applicable.  For this reaction, the only such diagram is Fig. \ref{fig:strongia}b where the 3-momentum in the nucleon propagator is large ($\sim\tilde{p}$).  For that propagator, we use the corrected version,
\be
D_N(p)=\frac{i}{p^0-\vec{p}\;^2/2M_N+i\epsilon}\approx-\frac{i}{m_\pi}.
\ee
Using this propagator rather than the $\nu=0$ version doubles the size of Fig. \ref{fig:strongia}b.  Nevertheless, this diagram (minus its $\omega_\pi=0$ analog) is already very small.  Thus the net effect of correcting the propagators is small for this reaction at this order.

\begin{figure}
\centering
\includegraphics[height=2in]{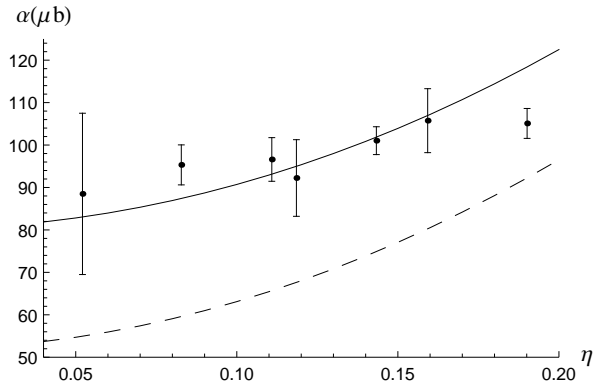}
\caption{Cross section for $np\rightarrow d\pi^0$ in terms of $\alpha=\sigma/\eta$ as a function of the pion center of mass momentum, $\eta=q/m_{\pi^0}$.  Circles with error bars display the experimental results of Ref. \cite{Hutcheon:1989bt}.  The dashed line displays the results of including the diagrams in Fig. \ref{fig:stronglo} and the solid line displays the results of also including the recoil terms discussed in the text.\label{fig:alpha}}
\end{figure}

Including all these recoil corrections (especially Fig. \ref{fig:strongrecoil}a) brings the theoretical cross section near the experimental results as shown by the solid curve of Fig. \ref{fig:alpha} \cite{Hutcheon:1989bt}.
Due to the relative scatter of the data shown in Fig. \ref{fig:alpha} it is difficult to tell how well the theory is reproducing the experiment.  Regardless, it is clear that theoretical improvement has been made.

\begin{figure}
\centering
\includegraphics[height=1in]{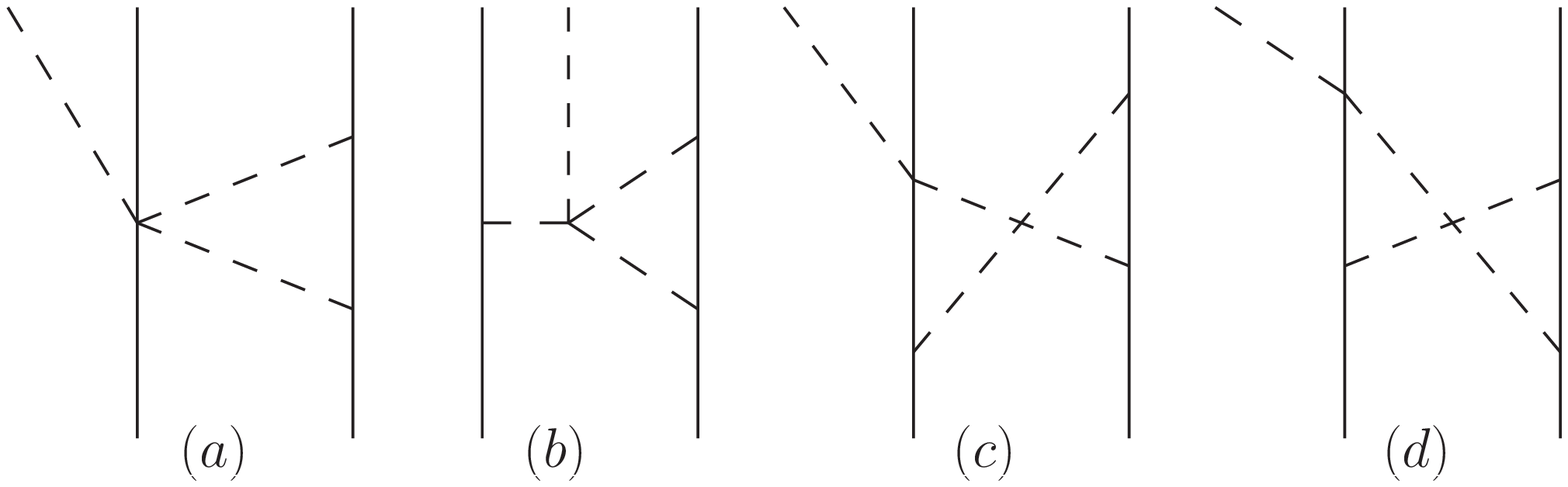}
\caption{Irreducible loops.  Solid lines represent nucleons and dashed lines represent pions.\label{fig:irrloops}}
\end{figure}

\begin{figure}
\centering
\includegraphics[height=1in]{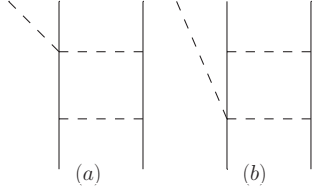}
\caption{Reducible loops.  Solid lines represent nucleons and dashed lines represent pions.\label{fig:redloops}}
\end{figure}

It should also be noted that the subtlety of reducibility and recoil corrections in this reaction resolves questions about NLO loop diagrams discovered by Ref. \cite{Gardestig:2005sn}.  Namely, the sum of all the NLO irreducible loops in Fig. \ref{fig:irrloops} is found to be proportional to $\vec{p}$.  This is a problem because such sensitivity of the operator to the NN wavefunction is not physical.
Again, we consider including OPE in the operator, this time for the rescattering diagram.  There are two resulting diagrams shown in Fig. \ref{fig:redloops}.  
It was shown in Ref. \cite{Lensky:2005jc} that in these reducible loops, the recoil corrections to the nucleon propagators need to be included in addition to the WT's recoil correction.  Ref. \cite{Lensky:2005jc} then showed that (part of) the energy dependence of the WT vertex \lqt cancels" one of the nucleon propagators leaving a reducible diagram similar to Fig. \ref{fig:irrloops}a.  This diagram is equal in magnitude and opposite in sign to the aforementioned NLO sum, resolving the issue.  The other term that remains from the original loop integral after this manipulation is still of reducible form but now has an on-shell WT vertex $\sim2m_\pi$.  This term would already appear upon convolution of the rescattering diagrams discussed above (including recoil corrections) with external wavefunctions, i.e. this term is truly reducible.  This result can be stated another way: cancellation of the irreducible loops comes from a short range piece of including OPE in the rescattering operator.  Thus a complete NLO calculation must include this OPE.  However, its remaining reducible piece would only be included in the calculation if the recoil correction to the WT vertex is included in the rescattering operator.  Therefore it is both consistent and necessary to include the WT recoil correction which was shown above to reproduce the total cross section.

\subsection{P-Wave Observables\label{sec:pwave}}

Another important test of the theory is how well it describes p-wave pions \cite{Baru:2009fm}.  This is especially important for the asymmetry, which involves strong p-waves at leading order.  The differential cross section can be expanded in Legendre polynomials,
\be
\frac{d\sigma}{d\Omega}=\alpha_0+\alpha_1P_1(\cos(\theta))+\alpha_2P_2(\cos(\theta))+...,\label{eq:legendre}
\ee
where $\theta$ is the angle between $\vec{p}$ and $\vec{q}$.  Note that the total cross section plotted in Fig. \ref{fig:alpha} is $\alpha=4\pi\alpha_0/\eta$.  As discussed in Appendix \ref{sec:observables}, $\alpha_2$ receives contributions almost exclusively from p-wave pions.  The ratio $\alpha_2/\alpha_0$ is therefore used as a test for this part of the theory.  We find that the diagrams of Fig. \ref{fig:stronglo} along with their recoil corrections overestimate the data by approximately a factor of two.  Upon closer inspection we find that the $^1S_0$ amplitude (which is known to be small) is relatively large.  This amplitude is coming mainly from the Delta diagram, as can be seen in Appendix \ref{sec:observables} where the values of the reduced matrix elements are listed.  To remedy the situation in the simplest way possible, we implement a cutoff for the Delta diagram
\bea
D_\pi=\frac{-i}{\vec{p}\,^2+\mu^2}\to D_\pi^c(\Lambda)&\equiv&\frac{-i}{\vec{p}\,^2+\mu^2}\left(\frac{\Lambda^2}{\vec{p}\,^2+\Lambda^2}\right)\nonumber
\\
&=&\left(\frac{-i}{\vec{p}\,^2+\mu^2}-\frac{-i}{\vec{p}\,^2+\Lambda^2}\right)\frac{\Lambda^2}{\Lambda^2-\mu^2}.\label{eq:cutoff}
\eea
One can show that doing this essentially softens the OPE potential for $r<\log(\Lambda/\mu)/\Lambda$.  Clearly this is not an acceptable long-term solution for an effective field theory, but the fact that such a procedure is necessary is interesting given that the reaction occurs at an energy  $\sim150\ \text{MeV}$ below the Delta resonance.  That p-wave pion production is highly sensitive to the strength of its contact term was discussed by Ref. \cite{Hanhart:2000gp}.

\begin{figure}
\centering
\includegraphics[height=2in]{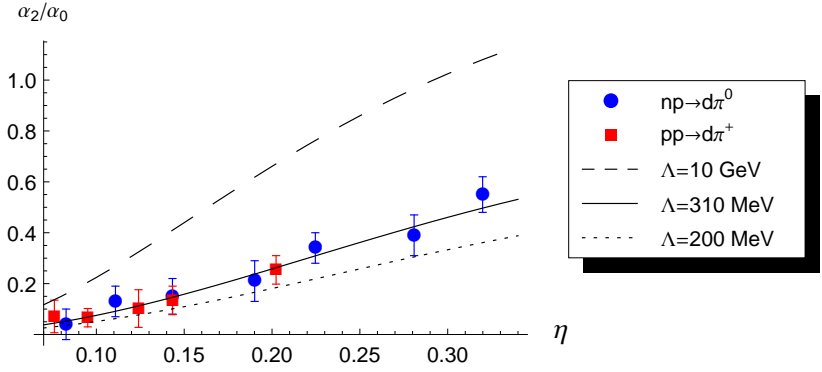}
\caption{\label{fig:a2}Legendre coefficients of the differential cross section for different values of the cutoff.  Data is from an $np\to d\pi^0$ experiment (Ref. \cite{Hutcheon:1989bt}, circles) and an $\vec{p}p\to d\pi^+$ experiment (Ref. \cite{Heimberg:1996be}, squares) in which the data have not be corrected for Coulomb effects.}
\end{figure}

Figure \ref{fig:a2} shows the effects of this cutoff on the ratio $\alpha_2/\alpha_0$, where $\Lambda=10\ \text{GeV}$ represents the original theory (such a large cutoff has no significant effect).
Note that the amplitudes for $np\to d\pi^0$ are related to those for $pp\to d\pi^+$ (which are bigger by $\sqrt{2}$) when charge symmetry is respected.  Thus the ratio plotted should have the same value for both reactions.  Also note that for simplicity we used a single cutoff, and that if we had used one at each vertex of the OPE we would have found $\Lambda\to\sqrt{2}\Lambda$.  By adjusting the cutoff to fit the data, we find $\Lambda=310\ \text{MeV}$.

\begin{figure}
\centering
\includegraphics[height=2in]{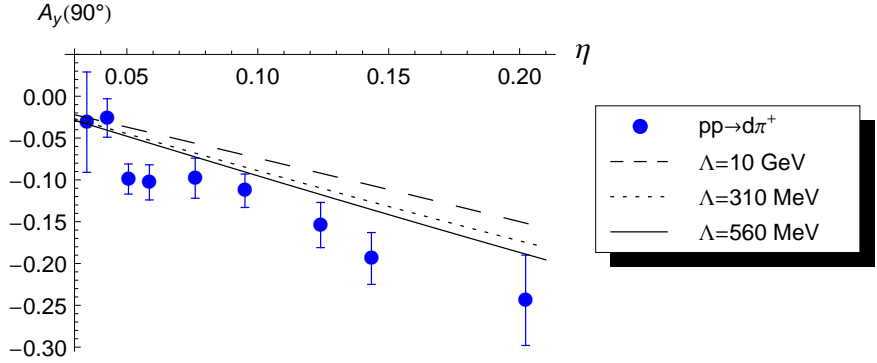}
\caption{\label{fig:ay}Analyzing power for different values of the cutoff.  Data is from a $\vec{p}p\to d\pi^+$ experiment (Ref. \cite{Heimberg:1996be}, circles) in which the data have not be corrected for Coulomb effects.}
\end{figure}

Another useful observable for testing p-wave pion production is the analyzing power, $A_y$, which is defined 
\bea
A_y(\theta)&\equiv&\frac{d\sigma_\uparrow(\theta)-d\sigma_\downarrow(\theta)}{d\sigma_\uparrow(\theta)+d\sigma_\downarrow(\theta)}\label{eq:aydef}
\\
d\sigma_{\uparrow,\downarrow}(\theta)&\equiv&\frac{|\vec{q}|}{64\pi^2s|\vec{p}|}\frac{1}{4}\sum_{m_d,m_2}\left|\mathcal{M}\left(m_{1,y}=\pm1/2,\theta\right)\right|^2,\label{eq:dsdoup}
\eea
where $m_{1,y}=\pm1/2$ refers to the fact that the beam is polarized perpendicular to the scattering plane.  In the z-basis, these states are
\be
\mathcal{M}\left(m_{1,y}=\pm1/2\right)=\frac{\mathcal{M}\left(m_1=1/2\right)\pm i\mathcal{M}\left(m_1=-1/2\right)}{\sqrt{2}}.
\ee
As shown in Appendix \ref{sec:observables}, $A_y$ is proportional to the product of s-wave and p-wave amplitudes.  Figure \ref{fig:ay} shows the effects of the cutoff on this observable.
Again, charge symmetry implies that $A_y$ should be the same for both neutral and charged pion production.  We find the best agreement with the data for $\Lambda=560\ \text{MeV}$.  Below, we will display our results using both the original theory and a cutoff taken to be the geometric mean of two fits, $\Lambda=417\ \text{MeV}$.

\section{Charge Symmetry Breaking\label{sec:csb}}

\begin{figure}
\centering
\includegraphics[height=1in]{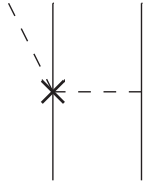}
\caption{Leading CSB contribution.  Solid lines represent nucleons, dashed lines represent pions, and crosses represent $\nu=1$ CSB vertices.\label{fig:csbrs}}
\end{figure}

The fact that the up and down quarks have different mass is reflected in the Lagrangian by including terms which break chiral symmetry \cite{Weinberg:1994tu}.  The leading such terms are given in Eq. (\ref{eq:l1}) and have coupling constants $\delta m_N$ and $\overline{\delta}m_n$ which are constrained by
\be
\delta m_N+\overline{\delta}m_N=M_n-M_p.
\label{eq:csbsize}
\ee
The $\delta m_N$ term has its origins in the quark mass difference and its size is $\sim (m_d-m_u)\equiv\epsilon(m_d+m_u)$ with $\epsilon\approx1/3$.  The formalism of $\chi$PT tells us that $(m_d+m_u)\propto m_\pi^2$, and so dimensional analysis along with the natural QCD scale, $M_N$, yields $\delta m_N\sim\epsilon m_\pi^2/M_N$.  The $\overline{\delta}m_N$ term is of electromagnetic origins, but is of the same order as $\delta m_N$.  These CSB operators appear in the rescattering diagram depicted in Fig. \ref{fig:csbrs}
where the CSB vertex is denoted with a cross.  The size of this diagram is $\delta m_N/\tilde{p}\approx\epsilon m_\pi^2/(M_N\tilde{p})=\epsilon\chi^3$.  Note that although the full nucleon mass difference appears explicitly in the Lagrangian at this order, the corresponding operator ($N^\dagger\tau_3N$) does not change the parity and thus does not contribute to an asymmetry.  In Sec. \ref{sec:discussion}, we will discuss another source of CSB coming from a more detailed evaluation of the diagram of Fig. \ref{fig:stronglo}b which was made by Ref. \cite{Filin:2009yh}.

\begin{figure}
\centering
\includegraphics[height=1in]{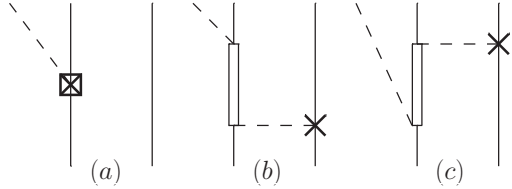}
\caption{$\nu=2$ CSB contributions.  Solid lines represent nucleons, double solid lines represent $\Delta$'s, dashed lines represent pions, crosses represent $\nu=2$ CSB vertices, and the boxed cross represents the full impulse CSB diagram including OPE.\label{fig:csbpwave}}
\end{figure}

Another CSB term given in Eq. (\ref{eq:l2}) involves one derivative and one $m_\pi^2$ ($\beta_1\sim\epsilon m_\pi^2/M_N^2$) and is thus a $\nu=2$ vertex with momentum dependence $|\vec{q}\,|$.  This vertex appears in the diagrams of Fig. \ref{fig:csbpwave} whose sizes are $\beta_1q/m_\pi\approx\epsilon\eta\chi^4$.  In Fig. \ref{fig:csbpwave}a, the boxed cross represents the sum analogous to Fig. \ref{fig:strongiasum} for the CSB impulse diagram.  Again, the wavefunction corrections are small ($2\%$).

\begin{figure}
\centering
\includegraphics[height=3in]{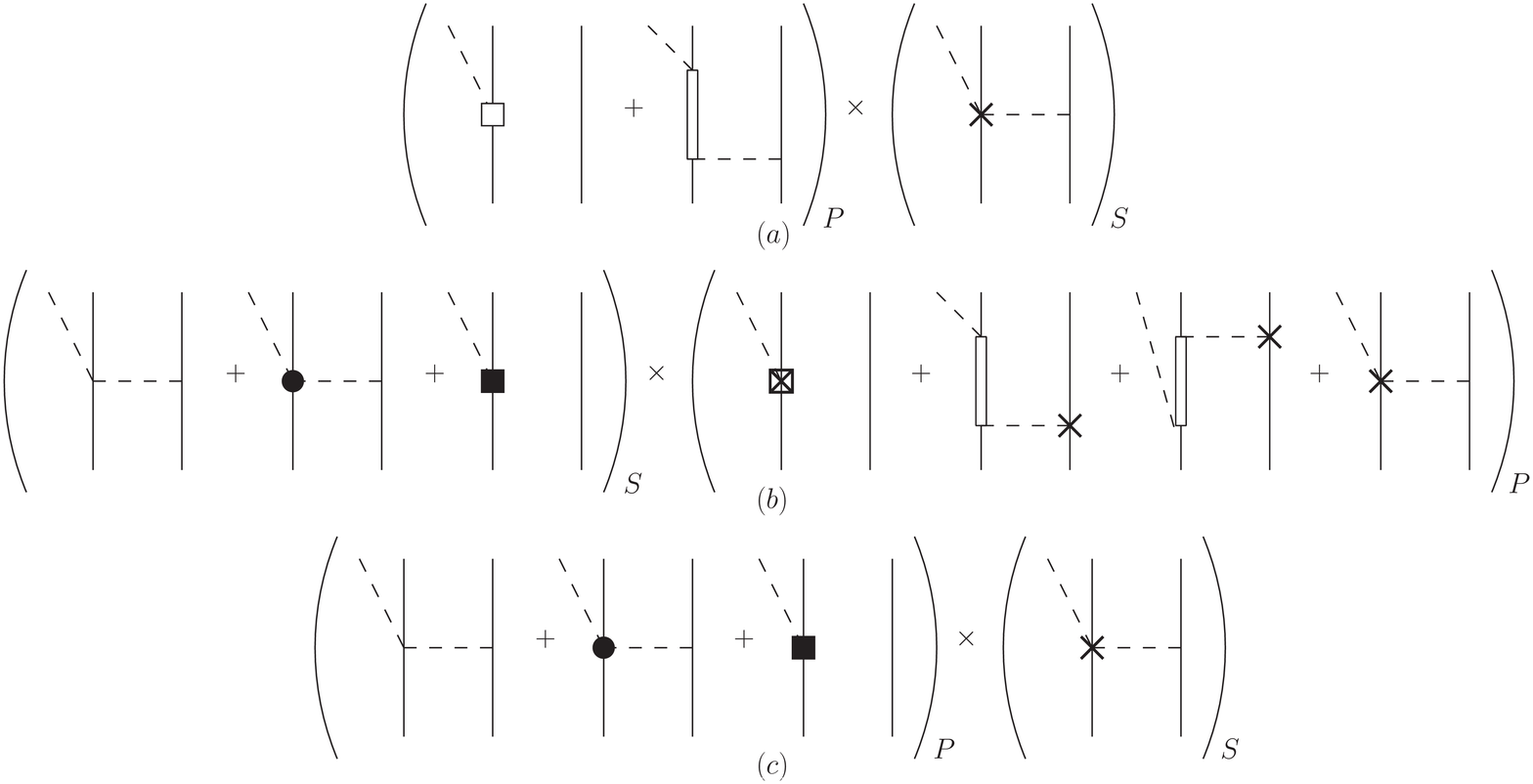}
\caption{Interference terms for the asymmetry in $np\rightarrow d\pi^0$.  Solid lines represent nucleons, double solid lines represent $\Delta$'s, and dashed lines represent pions.  The filled circle and square represent $\nu=1$  strong vertices and crosses represent CSB vertices.  The boxes represent full impulse diagrams in the sense of Fig. \ref{fig:strongiasum}.\label{fig:asymmetry}}
\end{figure}

As mentioned in Sec. \ref{sec:kinematics}, contributions to the asymmetry come from interference terms between the p-wave part of the strong amplitude and the s-wave part of the CSB amplitude, and vice versa.  The issue is somewhat complicated because, in contrast to threshold emission, each diagram can contribute in both the s-wave and the p-wave.  However, contributions to the sub-leading parity (s-wave for the impulse and Delta diagrams and p-wave for the rescattering diagrams) are formally higher order.  For example, the strong rescattering diagram for p-wave pions comes with a $m_\pi$ from the WT vertex, a $1/\tilde{p}\,^2$ from the pion propagator, and a $q$ from the pion emission in the p-wave.  Thus the diagram counts as $\sim\eta\chi^2$.  The contributions to the asymmetry are depicted in Fig. \ref{fig:asymmetry}.
Fig. \ref{fig:asymmetry}a includes strong p-waves and CSB s-waves and has size $(\eta)\times(\epsilon\chi^3)$.  Fig. \ref{fig:asymmetry}b includes strong s-waves and CSB p-waves and has size $(\chi)\times(\epsilon\eta\chi^4)$.  Fig. \ref{fig:asymmetry}c includes strong p-waves and CSB s-waves and has size $(\eta\chi^2)\times(\epsilon\chi^3)$.  Thus we find that in these kinematics Fig. \ref{fig:asymmetry}a ($\sim\epsilon\eta\chi^3$) is the LO contribution, and Figures \ref{fig:asymmetry}b,c ($\sim\epsilon\eta\chi^5$) both come in at NLO.  Other interference terms involving these diagrams are higher order, $\sim\epsilon\eta^3\chi^5$ or smaller.  Finally we note that this work is not intended to be a complete NLO calculation as loops and higher order vertices may come in at NLO.

\section{CSB Results\label{sec:results}}

In this section we discuss the CSB results of our calculation.  The coupling of the WT vertex (and its recoil correction) is determined by chiral symmetry and we use $f_\pi=91.9$ MeV.  For the impulse vertices we use the values $g_A=1.267$ and $h_A=2.1g_A$.  We use the following masses: $M_N=(M_n+M_p)/2=938.919$ MeV, $M_n-M_p=1.293$ MeV, $M_\Delta=1232$ MeV, and $m_\pi=m_{\pi^0}=134.977$ MeV.  Consider now the CSB coupling constants.  The Cottingham formula can be used to obtain $\overline{\delta}m_N=-(0.76\pm0.30)$ MeV \cite{Gasser:1982ap}.  The constraint of Eq. (\ref{eq:csbsize}) then fixes $\delta m_N=2.05\pm0.30$ MeV.  The combination of these parameters that appears in the asymmetry is $\delta m_N-\overline{\delta}m_N/2=2.4\pm0.3$ MeV.  It is also noted in Ref. \cite{vanKolck:2000ip} that other models predict different values for $\overline{\delta}m_N$ leading to $1.83\leq\frac{\delta m_N}{\text{MeV}}\leq2.83$.

Even less is known about $\beta_1$, the impulse CSB coupling.  As a starting point, Ref. \cite{vanKolck:1996rm} notes that this CSB operator can be viewed as arising from $\pi-\eta$ mixing.  The result shown is that
\be
\beta_1=\frac{g_\eta f_\pi}{M_Nm_\eta^2}\langle\pi^0|H|\eta\rangle=c_\eta\left(\frac{\epsilon\ m_\pi^2}{M_N^2}\right).
\label{eq:beta1}
\ee
As discussed in the review \cite{Miller:2006tv},
\be
0.10\leq\frac{g_\eta^2}{4\pi}\leq0.51.
\label{eq:etasize}
\ee
Also, Ref. \cite{Coon:1995xp} gives $\langle\pi^0|H|\eta\rangle=-0.0039\ \text{GeV}^2$, and we use $m_\eta=547.51$ MeV.  These values result in $-0.47\leq c_\eta\leq-0.21$.  Thus it is at least plausible that the $\beta_1$ term could originate naturally from $\eta-\pi$ mixing.  We note that the $\eta'$ could also give such a term, but do not consider it here.  Using Eqs. (\ref{eq:beta1}) and (\ref{eq:etasize}), we obtain $-3.2\times10^{-3}\leq\beta_1\leq-1.4\times10^{-3}$.  Note that the value used in the original calculation of the asymmetry by Ref. \cite{Niskanen:1998yi} was $\beta_1=-8.7\times10^{-3}$, which we refer to as the \lqt extreme value".  However, recall from the above discussion that the natural size of the p-wave CSB is $\beta_1\sim-\frac{\epsilon m_\pi^2}{M_N^2}\approx-6\times10^{-3}$.  Thus even though its origins may not lie exclusively with the $\eta$, the aforementioned \lqt extreme" value for $\beta_1$ is not extreme at all from the effective field theory's point of view.

In Table \ref{tab:results}, we display our results which are obtained by using Eq. (\ref{eq:csbsize}) to eliminate $\overline{\delta}m_N$ so that the diagram of Fig. \ref{fig:csbrs} is proportional to
\be
\delta m_N-\frac{\overline{\delta} m_N}{2}=\frac{3\delta m_N}{2}-\frac{M_n-M_p}{2}.
\ee
Because the asymmetry is linear in the CSB amplitudes (and therefore the CSB parameters), we are able to present our results as a set of coefficients, $\{x,y,z\}$ defined by
\be
A_{fb}\times10^4=x\cdot\left(\frac{\delta m_N}{\text{MeV}}\right)+y\cdot(\beta_1\times10^3)+z.
\label{eq:parameterization}
\ee
The primary advance made in this work over the the previous calculation of Ref. \cite{vanKolck:2000ip} (including the rescattering and impulse recoil corrections) is shown in moving from the top four rows to the next four rows.  At LO this simply increases $\alpha_0$, but at NLO it affects both the numerator and the denominator of the asymmetry.
\begin{table}
\caption{Asymmetry in $np\rightarrow d\pi^0$ as a function of CSB parameters $\delta m_N$ and $\beta_1$.  \lqt LO" and \lqt NLO" represent the sums discussed in Fig. \ref{fig:asymmetry}.  The term \lqt z" arises from the influence of the term proportional to $\frac{M_n-M_p}{2}$.\label{tab:results}}
\begin{center}
\renewcommand{\tabcolsep}{3mm}
\begin{tabular}{|c|c|c|c|c|c|}
\hline
\multicolumn{5}{|c|}{$A_{fb}\times10^4=x\cdot(\frac{\delta m_N}{\text{Mev}})+y\cdot(\beta_1\times10^3)+z$} & $A_{fb}(\delta m_N=2.05\ \text{MeV},$\\ \cline{1-5}
Order & Delta Cutoff & x & y & z & $\beta_1=-3.2\times10^{-3})\times10^4$\\ \noalign{\hrule height 1.5pt}
LO (no recoil) & None & 33.7 & 0 & -14.5 & 54.6\\ \hline
LO (no recoil) & $\Lambda=417\ \text{MeV}$ & 27.6 & 0 & -11.9 & 44.7\\ \hline
NLO (no recoil) & None & 37.6 & 1.4 & -16.2 & 56.4\\ \hline
NLO (no recoil) & $\Lambda=417\ \text{MeV}$ & 32.5 & 1.8 & -14.0 & 47.0\\ \hline \noalign{\hrule height 1.5pt}
LO & None & 25.0 & 0 & -10.8 & 40.4\\ \hline
LO & $\Lambda=417\ \text{MeV}$ & 18.9 & 0 & -8.2 & 30.7\\ \hline
NLO & None & 28.1 & 1.4 & -12.1& 41.0\\ \hline
NLO & $\Lambda=417\ \text{MeV}$ & 22.1 & 1.6 & -9.5 & 30.7\\ \noalign{\hrule height 1.5pt}
NLO & $\Lambda=310\ \text{MeV}$ & 20.1 & 1.7 & -8.7 & 27.3\\ \hline
NLO & $\Lambda=560\ \text{MeV}$ & 24.0 & 1.6 & -10.4 & 33.9\\ \hline
\end{tabular}
\end{center}
\end{table}

The experiment of Ref. \cite{Opper:2003sb} found $A_{fb}=[17.2\pm9.7]\times10^{-4}$.  The first calculation of the asymmetry used a $N\Delta$ coupled channel formalism and included the CSB impulse vertex as well as other, smaller effects arising directly from the neutron-proton mass difference \cite{Niskanen:1998yi}.  This study reported $A_{fb}=-28\times10^{-4}$.  The second calculation included only the CSB rescattering vertex, and found $A_{fb}=60\times10^{-4}$ \cite{vanKolck:2000ip}.  Both these calculations were preformed before the work of Ref. \cite{Lensky:2005jc} which brought the total cross section into agreement with experiment.  Our work brings the asymmetry calculation up to date.

As shown in Table \ref{tab:results}, the value of $A_{fb}$ is over-predicted for the set of physically reasonable parameters used in the last column.  Nevertheless, for the most \lqt extreme" set of parameters discussed above ($\delta m_N=1.83\ \text{MeV}$, $\beta_1=-8.7\times10^{-3}$) the cutoff NLO calculation yields $A_{fb}=16.9\times10^{-4}$.  The effects of using different values for the cutoff can be seen in the last two rows of the table.

\section{Discussion\label{sec:discussion}}

We have mentioned that it is difficult to tell how well the current theory is reproducing the total cross section.  We have also seen that there is reason for concern regarding the theoretical description of p-wave pions, which comprise the entire strong contribution to the LO asymmetry.  Because the total cross section is dominated by the rescattering diagram, small changes to the p-wave amplitudes are able to significantly modify the asymmetry while only slightly changing the total cross section.  As a temporary solution, we implemented a cutoff in the Delta diagram and thereby achieved acceptable agreement with the p-wave data.  Another solution to this problem is to use a coupled-channel $N\Delta$ potential for the initial state.  This approach was taken by Ref. \cite{Filin:2009yh} who were able to achieve good fits to these data without a cutoff, since the OPE of the Delta diagram is then part of the wavefunction.

There are other significant differences between Ref. \cite{Filin:2009yh} and this work.  First of all, they discovered an additional CSB contribution to the asymmetry which is a consequence of the time derivative in the WT vertex.  The effect of this contribution is equivalent to a change in the CSB rescattering diagram,
\be
\frac{3\delta m_N}{2}-\frac{M_n-M_p}{2}\to\frac{3\delta m_N}{2}.
\label{eq:newcsb}
\ee
Thus in order to update our calculation, we drop the fifth column of Table \ref{tab:results} (the \lqt z" column).  Secondly, they used experimental data (from pionic deuterium) to determine the Legendre coefficient, $\alpha_0=1.93\ \mu\text{b}$.  This is significantly larger than the theoretical value we use, $\alpha_0=1.49\ \mu\text{b}$ at LO ($\alpha_0=1.28\ \mu\text{b}$ for $\Lambda=417\ \text{MeV}$), and leads to a smaller value for the for the asymmetry.  Note that experiments for neutral \cite{Hutcheon:1989bt} and charged \cite{Heimberg:1996be} pion production found $\alpha_0=1.39\ \mu\text{b}$ and $\alpha_0=1.64\ \mu\text{b}$ (Coulomb corrected) respectively.  Finally, they do not include the NLO CSB diagrams.

For the sake of comparison, we used our code to calculate the LO asymmetry with $\Lambda=417\ \text{MeV}$, without the \lqt z" column, using $\alpha_0=1.93$, and using their quoted values for $g_A$ and $f_\pi$.  For these choices we obtain $A_{fb}=14.0\frac{\delta m_N}{\text{MeV}}\times10^{-4}$, which is to be compared with their result of $A_{fb}=11.5\frac{\delta m_N}{\text{MeV}}\times10^{-4}$.  They use the Cottingham sum rule to obtain $\delta m_N=2.0\ \text{MeV}$ and thus $A_{fb}=23\times10^{-4}$, which only overestimates the data by $0.6\sigma$.  Finally, we display our best result without the \lqt z" column,
\be
A_{fb}=22.1\,\frac{\delta m_N}{\text{MeV}}+1.6\,(\beta_1\times10^3).
\ee
For the set of parameters used in Table \ref{tab:results}, $A_{fb}=40.2\times10^{-4}$ which is an overestimation of the data by $2.4\ \sigma$.

Several issues remain to be understood theoretically.  Firstly, it appears that a large contact term will be required to suppress the $^1S_0$ channel in the strong amplitude if one uses a purely NN initial state.  The interesting physics observation here is that the Delta part of the NN wavefunction seems to be much more active than it is NN scattering.  Secondly, the experimental cross section used by Ref. \cite{Filin:2009yh} is not well predicted by theory at NLO, and this fact plays a large role in the overprediction of our calculation.  This situation becomes even worse when a cutoff is used to decrease the p-wave amplitudes.  Finally, the existence of multiple mass scales greatly complicates the power counting for this reaction, and it is clear that a converging expansion cannot yet be definitively claimed.  For these reasons, we conclude that further calculations are necessary.  In particular, one should extend the calculation to next order while examining both the power counting of recoil terms and the reducibility of loops.

\begin{acknowledgments}
We would like to thank U. van Kolck, D. Phillips, and the authors of Ref. \cite{Filin:2009yh} for useful discussions.  This research was supported in part by the U.S. Department of Energy.
\end{acknowledgments}

\appendix

\section{Lagrange Densities\label{sec:lagrangian}}

The $\nu=0$ lagrangian of HB$\chi$PT (with isovectors in $\mathbf{bold}$ font) with the Delta included as an explicit degree of freedom is \cite{Cohen:1995cc}
\bea
\mathcal{L}^{\left(0\right)}&=&\frac{1}{2}\left(\partial\bpi\right)^2-\frac{1}{2}m_\pi^2\bpi^2+N^\dagger i\partial_0N\nonumber
\\
&-&\frac{1}{4f_\pi^2}N^\dagger\left(\btau\cdot\left(\bpi\times\dot{\bpi}\right)\right)N+\frac{g_A}{2f_\pi}N^\dagger\left(\btau\cdot\vec{\sigma}\cdot\vec{\nabla}\bpi\right)N\nonumber
\\&+&\Delta^\dagger\left(i\partial_0-\delta\right)\Delta+\frac{h_A}{2f_\pi}\left[N^\dagger\left(\mathbf{T}\cdot\vec{S}\cdot\vec{\nabla}\bpi\right)\Delta+h.c.\right]+...,
\label{eq:l0}
\eea
where $\mathbf{T}$ and $\vec{S}$ are the $3/2\rightarrow1/2$ isospin and spin transition operators, and $\btau$ and $\vec{\sigma}$ are the pauli matrices acting on the isospin and spin of a single nucleon.  The \lqt$+...$" indicates that only the terms used in this calculation are shown.

The $\nu=1$ lagrangian includes propagator corrections, recoil terms, and the leading s-wave CSB operator
\bea
\mathcal{L}^{\left(1\right)}&=&\frac{1}{2m_N}N^\dagger\nabla^2N+\frac{1}{2m_N}\left[\frac{1}{4f_\pi^2}iN^\dagger\btau\cdot\left(\bpi\times\vec{\nabla}\bpi\right)\cdot\vec{\nabla}N-\frac{g_A}{2f_\pi}iN^\dagger\btau\cdot\dot{\bpi}\vec{\sigma}\cdot\vec{\nabla}N+h.c.\right]\nonumber
\\
&+&\frac{1}{2m_N}\Delta^\dagger\nabla^2\Delta-\frac{1}{2m_N}\frac{2h_A}{2f_\pi}\left[iN^\dagger\mathbf{T}\cdot\dot{\bpi}\vec{S}\cdot\vec{\nabla}\Delta+h.c.\right]\nonumber
\\
&+&\frac{\delta m_N}{2}N^\dagger\left[\tau_3-\frac{2}{4f_\pi^2}\pi_3\btau\cdot\bpi\right]N+\frac{\overline{\delta} m_N}{2}N^\dagger\left[\tau_3+\frac{2}{4f_\pi^2}\left(\pi_3\btau\cdot\bpi-\tau_3\bpi^2\right)\right]N+...\,.\label{eq:l1}
\eea

Although there are a host of $\nu=2$ terms, we list the just the CSB term relevant for this calculation
\be
\mathcal{L}^{\left(2\right)}=\frac{\beta_1}{2f_\pi}N^\dagger\vec{\sigma}\cdot\vec{\nabla}\pi_3N+...\,.
\label{eq:l2}
\ee

\section{Defining Reduced Matrix Elements\label{sec:reduced}}

One can show that for s-wave pions, the production operator is always either a scalar or a rank-two tensor while for p-wave pions, it is always a rank-one tensor.  This guides the following definition of the reduced matrix elements.  Note that the Clebsch-Gordan coefficients which will be summed over as well as the Spherical Harmonics describing the angular distribution of the differential cross section are \lqt pulled out."  First we define the strong reduced matrix elements,
\bea
\left\la f\mid\hat{\mathcal{M}}^{str}_{l_\pi=0}\mid i\right\ra&=&\left(\frac{1}{\sqrt{3}}A_0+\frac{\left\la1\,m_f,2\,0\mid1\,m_f\right\ra}{\sqrt{3}}A_2\right)e^{i\delta_1}\text{cg}_1\nonumber
\\
&\times&\left\la1\,m_s,1\,m_f-m_s\mid1\,m_f\right\ra Y^{1\,*}_{m_f-m_s}(\hat{p})\label{eq:redstrs}
\\[0.1in]
\left\la f\mid\hat{\mathcal{M}}^{str}_{l_\pi=1}\mid i\right\ra&=&\frac{1}{\sqrt{3}}Be^{i\delta_0}\text{cg}_0\delta_{m_f,0}Y^0_0(\hat{p})+\frac{\left\la2\,m_f,1\,0\mid1\,m_f\right\ra}{\sqrt{3}}Ce^{i\delta_2}\text{cg}_0Y^{2\,*}_{m_f}(\hat{p}),\label{eq:redstrp}
\eea
where $m_s=m_1+m_2$, $\text{cg}_0=\cgz$, and $\text{cg}_1=\left\la1/2\,m_1,1/2\,m_2\mid1\,m_s\right\ra$.  $A_0$ and $A_2$ are the s-wave reduced matrix elements, and $B$ and $C$ are the p-wave reduced matrix elements.  To clarify the notation consider $A_2$, for example.
\be
A_2=\int dr\,r^2\left(\frac{u_d(r)}{r}\la(10)1\mid\mid+\frac{w_d(r)}{r}\la(12)1\mid\mid\right)\hat{\mathcal{M}}^{str}_{l_\pi=0,J=2}\left(4\pi\,i\frac{u_{1,1}(r)}{pr}\mid\mid(11)1\ra\right),
\ee
where the subscript $J=2$ on the $\hat{\mathcal{M}}$ indicates that we are using the portion of the operator that is a rank-two tensor in the space of total angular momentum.  Note also that we have used the following general definition of a reduced matrix element,
\be
\la(S'L')J'm_J'\mid T^k_q\mid(SL)Jm_J\ra\equiv\frac{\la Jm_J,kq\mid J'm_J'\ra}{\sqrt{2J'+1}}\la(S'L')J'\mid\mid T^k\mid\mid (SL)J\ra.
\ee
Similarly for the CSB reduced matrix elements,
\bea
\left\la f\mid\hat{\mathcal{M}}^{csb}_{l_\pi=0}\mid i\right\ra&=&\left(\frac{1}{\sqrt{3}}\overline{A}_0+\frac{\left\la1\,m_f,2\,0\mid1\,m_f\right\ra}{\sqrt{3}}\overline{A}_2\right)e^{i\overline{\delta}_1}\text{cg}_0Y^{1\,*}_{m_f}(\hat{p})\label{eq:redcsbs}
\\[0.1in]
\left\la f\mid\hat{\mathcal{M}}^{csb}_{l_\pi=1}\mid i\right\ra&=&\frac{\left\la1\,m_f,1\,0\mid1\,m_f\right\ra}{\sqrt{3}}\left(\overline{B}_\alpha e^{i\overline{\delta}_\alpha}+\overline{B}_\beta e^{i\overline{\delta}_\beta}\right)\text{cg}_1\delta_{m_f,m_s}Y^0_0(\hat{p})\nonumber
\\
&+&\frac{\left\la1\,m_f,1\,0\mid1\,m_f\right\ra}{\sqrt{3}}\left(\overline{C}_\alpha e^{i\overline{\delta}_\alpha}+\overline{C}_\beta e^{i\overline{\delta}_\beta}\right)\text{cg}_1\nonumber
\\
&&\qquad\times\left\la1\,m_s,2\,m_f-m_s\mid1\,m_f\right\ra Y^{2\,*}_{m_f-m_s}(\hat{p})\nonumber
\\
&+&\frac{\left\la2\,m_f,1\,0\mid1\,m_f\right\ra}{\sqrt{3}}\overline{D}e^{i\overline{\delta}_2}\text{cg}_1\left\la1\,m_s,2\,m_f-m_s\mid2\,m_f\right\ra Y^{2\,*}_{m_f-m_s}(\hat{p}),\label{eq:redcsbp}
\eea
where $\overline{A}_0$ and $\overline{A}_2$ are the s-wave reduced matrix elements and $\overline{B}$, $\overline{C}$, and $\overline{D}$ are the p-wave reduced matrix elements.  Also note that the strong phase shifts have been denoted $\delta_L$ for each of the three initial channels, the CSB phase shifts are denoted $\overline{\delta}_L$ for the $^1P_1$ and $^3D_2$ channels, and the coupled channel phase shifts are $\overline{\delta}_\alpha$ and $\overline{\delta}_\beta$.  Since the $^3S_1$ and $^3D_1$ channels are coupled, $\overline{B}$ and $\overline{C}$ are split into $\alpha$ and $\beta$ parts which have different phase shifts in the presence of initial state interactions.

Finally, for comparison purposes we include a translation between our reduced matrix elements and those of Ref. \cite{Baru:2009fm},
\bea
C_0&=&-\frac{1}{\sqrt{4\pi}}A_0e^{i\delta_0}\nonumber
\\
C_1&=&-i\frac{1}{\sqrt{6\pi}}Be^{i\delta_1}\nonumber
\\
C_2&=&i\sqrt{\frac{3}{4\pi}}Ce^{i\delta_2},
\eea
where the \lqt C's" are $pp\to d\pi^+$ amplitudes and isospin symmetry has been used to determine the translations.

\section{Diagram Technique\label{sec:method}}

\begin{figure}
\centering
\includegraphics[height=1.5in]{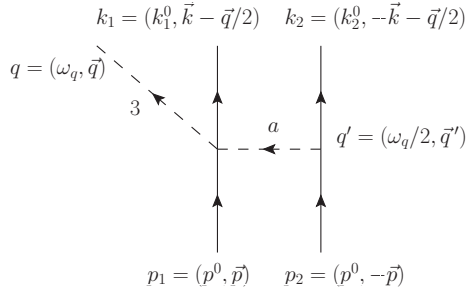}
\caption{Strong rescattering diagram.  Solid lines represent nucleons, dashed lines represent pions, and the pions' isospin z-components are 3 and a.\label{fig:strongrs}}
\end{figure}

To establish the diagram technique consider Fig. \ref{fig:strongrs} in light of Eq. (\ref{eq:mft}).
We write down the amplitude using momentum space Feynman rules, Fourier transform, and then convolve with the initial and final state wavefunctions.  The left line is taken to be \lqt nucleon 1" and we make the approximation that the exchanged pion carries half of the produced pion's energy, $q'=(\omega_q/2,\vec{q}\,')$.  Momentum conservation gives $\vec{q}\,'=\vec{k}-\vec{p}+\vec{q}/2$.  According to Eq. (\ref{eq:l0}) the WT vertex contributes $1/(4f_\pi^2)\,\tau_{1,b}\,\epsilon_{a3b}\,(\omega_q/2+\omega_q)$ while the $\pi NN$ vertex contributes $g_A/(2f_\pi)\,\tau_{2,a}\,\vec{\sigma}_2\cdot(-\vec{q}\,')$.  The momentum space propagator is $-i/(\vec{q}\,'^2+\mu^2)$ where $\mu^2=m_\pi^2-\omega_q^2/4$ is the effective mass of the rescattered pion.  Next, as discussed in Sec. \ref{sec:formalism}, we Fourier transform with respect to $\vec{l}=\vec{k}-\vec{p}$,
\bea
\int \frac{d^3l}{(2\pi)^3}\,e^{i\vec{l}\cdot\vec{r}}\frac{\vec{\sigma}_2\cdot\vec{q}\,'}{\vec{q}\,'^2+\mu^2}&=&e^{-i\vec{q}\cdot\vec{r}/2}\int \frac{d^3q'}{(2\pi)^3}\,e^{i\vec{q}\,'\cdot\vec{r}}\frac{\vec{\sigma}_2\cdot\vec{q}\,'}{\vec{q}\,'^2+\mu^2}\nonumber
\\
&=&e^{-i\vec{q}\cdot\vec{r}/2}\,\vec{\sigma}_2\cdot(-i\vec{\nabla})\,\frac{e^{-\mu r}}{4\pi r}\nonumber
\\
&=&\frac{i\mu}{4\pi}e^{-i\vec{q}\cdot\vec{r}/2}\,h(r)\,\vec{\sigma}_2\cdot\hat{r},\label{eq:rsft}
\eea
where $h(r)\equiv (1+1/\mu r)e^{-\mu r}/r$.

The deuteron has isospin 0 and the np wavefuction includes a $T=1$, $T_z=0$ isospinor $\left|1,0\right\ra$, and thus
\be
\la0,0\mid i\epsilon_{a3b}\tau_{1,b}\tau_{2,a}\mid1,0\ra=-2.
\ee
At this point, we have (defining $\hat{\mathcal{M}}'=\hat{\mathcal{M}}/\sqrt{2E_1\,2E_2\,2E_d}\,$)
\be
\left\la0,0\mid\hat{\mathcal{M}'_L}\mid1,0\right\ra=-i\frac{g_A}{2f_\pi}\,\frac{3\omega_q/2}{8\pi f_\pi^2}\,\mu\,h(r)\,e^{-i\vec{q}\cdot\vec{r}/2}\,\vec{\sigma}_2\cdot\hat{r}.
\label{eq:mstrrsl}
\ee
To calculate the diagram with the WT vertex on nucleon 2 we consider how each part of the left side of Eq. (\ref{eq:mstrrsl}) transforms under $1\leftrightarrow2$.  Since the strong part of the Lagrangian is invariant under isospin, $\hat{\mathcal{M}}$ is invariant.  The initial isospin ket, $\left|1,0\right\ra$ is invariant as well, but $\left|0,0\right\ra\rightarrow-\left|0,0\right\ra$.  Also note that $\vec{r}\rightarrow-\vec{r}$.  Thus,
\be
\left\la0,0\mid\hat{\mathcal{M}'_R}\mid1,0\right\ra=-i\frac{g_A}{2f_\pi}\,\frac{3\omega_q/2}{8\pi f_\pi^2}\,\mu\,h(r)\,e^{i\vec{q}\cdot\vec{r}/2}\,\vec{\sigma}_1\cdot\hat{r}.
\ee

Defining $\vec{S}\equiv(\vec{\sigma}_1+\vec{\sigma}_2)/2$, $\vec{\Sigma}\equiv(\vec{\sigma}_1-\vec{\sigma}_2)/2$, and
\bea
\mathcal{E}&\equiv&\text{exp}(i\vec{q}\cdot\vec{r}/2)+\text{exp}(-i\vec{q}\cdot\vec{r}/2)\nonumber
\\
\mathcal{O}&\equiv&\text{exp}(i\vec{q}\cdot\vec{r}/2)-\text{exp}(-i\vec{q}\cdot\vec{r}/2),\label{eq:eando}
\eea
we have the complete rescattering operator,
\be
\left\la0,0\mid\hat{\mathcal{M}}'_{RS}\mid1,0\right\ra=-i\,\gamma_{RS}\,h(r)\left(\mathcal{E}\,\vec{S}\cdot\hat{r}+\mathcal{O}\,\vec{\Sigma}\cdot\hat{r}\right),\label{eq:mstrrs}
\ee
where
\be
\gamma_{RS}\equiv\frac{g_A}{2f_\pi}\frac{3\omega_q/2}{8\pi f_\pi^2}\,\mu.
\ee

To proceed, we preform a partial wave expansion on $\mathcal{E}$ and $\mathcal{O}$ and just keep the leading term.  Note that we use the coordinate system defined by $\hat{q}=\hat{z}$.  Then we calculate the spin-angle matrix elements of the rank zero $\mathcal{E}\,\vec{S}\cdot\hat{r}$ and the rank one $\mathcal{O}\,\vec{\Sigma}\cdot\hat{r}$ operators.  The final expression for $\langle f\mid\hat{\mathcal{M}}\mid i\rangle$ simplifies to a radial integral which is computed numerically.  Note that the first term in Eq. (\ref{eq:mstrrs}) corresponds to s-wave pions because $\mathcal{E}$ carries $L=0$ while $\hat{r}$ carries $L=1$, and thus the operator will change the parity.  Likewise, the second term corresponds to p-wave pions.  In terms of the reduced matrix elements of Eqs. (\ref{eq:redstrs}) and (\ref{eq:redstrp}), we have
\bea
A_0^{RS}&=&\sqrt{2E_1\,2E_2\,2E_d}\,8\pi\,\gamma_{RS}\,\sqrt{2}\,K_1\nonumber
\\
A_2^{RS}&=&0\nonumber
\\
B^{RS}&=&\sqrt{2E_1\,2E_2\,2E_d}\,8\pi\,\gamma_{RS}\,\sqrt{3}\,K_0\nonumber
\\
C^{RS}&=&\sqrt{2E_1\,2E_2\,2E_d}\,8\pi\,\gamma_{RS}\,\sqrt{6}\,K_2,\label{eq:rsresults}
\eea
where the integrals are defined
\bea
K_1&\equiv&\int dr\,r^2\left(\frac{u_d(r)}{r}+\frac{w_d(r)}{\sqrt{2}r}\right)j_0\left(\frac{qr}{2}\right)\,h(r)\frac{u_{1,1}(r)}{pr}\nonumber
\\
K_0&\equiv&\int dr\,r^2\left(\frac{u_d(r)}{r}-2\frac{w_d(r)}{\sqrt{2}r}\right)j_1\left(\frac{qr}{2}\right)\,h(r)\frac{u_{0,0}(r)}{pr}\nonumber
\\
K_2&\equiv&\int dr\,r^2\left(\frac{u_d(r)}{r}-2\frac{w_d(r)}{\sqrt{2}r}\right)j_1\left(\frac{qr}{2}\right)\,h(r)\frac{u_{2,2}(r)}{pr}.\label{eq:rsintegrals}
\eea

\section{\label{sec:observables}Observables}

One experimental observable is the analyzing power, $A_y$, defined in Eq. (\ref{eq:aydef}).  In the strong sector, we find (neglecting $A_2$ which is numerically small)
\be
A_y=\frac{\sqrt{3}\cos(\phi)\sin(\theta)A_0\left(\sqrt{2}B\sin(\delta_0-\delta_1)+C\sin(\delta_2-\delta_1)\right)}{3A_0^2+B^2+C^2+\left(C^2-2\sqrt{2}BC\cos(\delta_2-\delta_0)\right)P_2(\cos\theta)},\label{eq:ay}
\ee
where the angular dependence is that of the nucleon relative momentum, $\vec{p}$, with respect to the pion momentum, $\hat{q}=\hat{z}$.  To compare with experimental results, which use $\hat{p}=\hat{z}$ and $\phi_\pi=0$, we need to set $\phi_N=\pi$ and $\theta_N=\theta_\pi$.

To calculate the differential cross section as well as the asymmetry, we need to square the sum of all the matrix elements, sum over $m_f$ and average over $m_1$ and $m_2$.  First we define
\be
\frac{1}{4}\sum\left|\left\la f\mid\hat{\mathcal{M}}^{tot}\mid i\right\ra\right|^2=M_0+M_1P_1(\cos\theta)+M_2P_2(\cos\theta)+M_3P_3(\cos\theta),
\ee
so that
\bea
\sigma&=&\frac{|\vec{q}\,|}{64\,\pi^2\,s\,|\vec{p}\,|}4\pi M_0\label{eq:sigmadef}
\\[0.1in]
A_{fb}&=&\frac{M_1-\frac{1}{4}M_3}{2M_0}.\label{eq:afbdef}
\eea
The results for the required quantities are
\bea
M_0&=&\frac{1}{48\,\pi}\left[3(A_0^2 + \overline{A}_0^2) + \frac{3}{5} (A_2^2 + \overline{A}_2^2) + \left(B^2 + \overline{B}_\alpha^2 + \overline{B}_\beta^2 + C^2 + \overline{C}_\alpha^2+\overline{C}_\beta^2 + \overline{D}^2\right)\right.\nonumber
\\
&&\qquad+\left.2\left(\overline{B}_\alpha\overline{B}_\beta + \overline{C}_\alpha\overline{C}_\beta\right)\cos\left(\overline{\delta}_\alpha-\overline{\delta}_\beta\right)\right]\label{eq:m0}
\\[.1in]
M_1&=&\frac{\sqrt{3}}{24\,\pi}\left[B\left(\overline{A}_0-2\sqrt{\frac{1}{10}}\overline{A}_2\right)\cos(\overline{\delta}_1-\delta_0)-\sqrt{2}C\left(\overline{A}_0-\frac{1}{5\sqrt{10}}\overline{A}_2\right)\cos(\overline{\delta}_1-\delta_2)\right.\nonumber
\\
&&\qquad+\left(A_0+\frac{1}{\sqrt{10}}A_2\right)\left(\left(\overline{B}_\alpha+\frac{1}{\sqrt{2}}\overline{C}_\alpha\right)\cos(\overline{\delta}_\alpha-\delta_1)+(\alpha\to\beta)\right)\nonumber
\\
&&\qquad-\left.\sqrt{\frac{3}{2}}\left(A_0-\frac{1}{5\sqrt{10}}A_2\right)\overline{D}\cos(\overline{\delta}_2-\delta_1)\right]\label{eq:m1}
\\[.1in]
M_2&=&\frac{1}{8\sqrt{10}\,\pi}\left[(A_0 A_2-2\overline{A}_0\overline{A}_2)-\frac{1}{2\sqrt{10}}(A_2^2-2\overline{A}_2^2)\right.\nonumber
\\
&&\qquad\qquad+\frac{5}{3\sqrt{10}}(C^2-\frac{1}{2}\overline{C}_\alpha^2-\frac{1}{2}\overline{C}_\beta^2+\frac{1}{2}\overline{D}^2-\overline{C}_\alpha\overline{C}_\beta\cos(\overline{\delta}_\alpha-\overline{\delta}_\beta))\nonumber
\\
&&\qquad\qquad-\frac{\sqrt{5}}{3}\left(2BC\cos(\delta_2-\delta_0)-\overline{B}_\alpha\overline{C}_\alpha-\overline{B}_\beta\overline{C}_\beta-(\overline{B}_\alpha\overline{C}_\beta+\overline{B}_\beta\overline{C}_\alpha)\cos(\overline{\delta}_\alpha-\overline{\delta}_\beta)\right)\nonumber
\\
&&\left.\qquad\qquad-\sqrt{\frac{5}{3}}\left(\overline{B}_\alpha+\frac{1}{\sqrt{2}}\overline{C}_\alpha\right)\overline{D}\cos(\overline{\delta}_2-\overline{\delta}_\alpha)-(\alpha\to\beta)\right]\label{eq:m2}
\\[.1in]
M_3&=&\frac{3}{40\,\pi}\sqrt{\frac{3}{5}}\left[C\overline{A}_2\cos(\overline{\delta}_1-\delta_2)-\frac{1}{\sqrt{3}}A_2\overline{D}\cos(\overline{\delta}_2-\delta_1)\right].\label{eq:m3}
\eea
Disregarding the small $M_3$ term, the asymmetry is proportional to Eq. (\ref{eq:m1}).  The physical content of Eq. (\ref{eq:m1})'s first line is the interference of strong p-wave pions with CSB s-wave pions and the second and third lines are strong s-wave and CSB p-wave.

Table \ref{tab:strmxels} shows the strong reduced matrix elements and Table \ref{tab:csbmxels} the CSB reduced matrix elements.  The CSB rescattering numbers were calculated including the new contributions discovered by Ref. \cite{Filin:2009yh}.  The $np$ phase shifts (in radians) which appear in the cross section according to Eqs. (\ref{eq:m0}-\ref{eq:m3}) are given in Tables \ref{tab:strongdeltas} and \ref{tab:csbdeltas}.

\begin{table}[htb]
\caption{\label{tab:strmxels}Strong reduced matrix elements}
\begin{center}
\renewcommand{\tabcolsep}{3mm}
\begin{tabular}{|c||c|c|c|c|}
\hline
Diagram & $A_0$ & $A_2$ & $B$ & $C$ \\ \noalign{\hrule height 1.5pt}
Impulse (w/ wfn corr) & 0 & 0 & -6.59 & 32.85 \\ \hline
Impulse Recoil & -5.62 & -0.21 & 1.17 & -8.56 \\ \hline
RS (w/ Recoil) & 81.12 & 0 & 0.54 & 1.66 \\ \hline
Delta (no cutoff) & 0 & 0 & -33.94 & 37.96 \\ \hline
Delta ($\Lambda=417\ \text{MeV}$) & 0 & 0 & -10.48 & 21.60 \\ \hline
\end{tabular}
\end{center}
\end{table}

\begin{table}[htb]
\caption{\label{tab:csbmxels}CSB reduced matrix elements}\begin{center}
\renewcommand{\tabcolsep}{3mm}
\begin{tabular}{|c||c|c|c|c|c|c|c|}
\hline
Diagram & $\overline{A}_0$ & $\overline{A}_2$ & $\overline{B}_\alpha$ & $\overline{B}_\beta$ & $\overline{C}_\alpha$ & $\overline{C}_\beta$ & $\overline{D}$ \\ \noalign{\hrule height 1.5pt}
Impulse ($\times\frac{1}{\beta_1}$) & 0 & 0 & 12.23 & 29.72 & -7.80 & -15.05 & -28.30 \\ \hline
RS ($\times\frac{100\,\text{MeV}}{\delta m_N^{str}}$) & -28.83 & 0 & -1.37 & 1.79 & 1.48 & 1.89 & -4.92 \\ \hline
Delta ($\times\frac{1}{\beta_1}$) & 0 & 0 & 12.60 & -7.71 & -8.47 & -2.86 & 22.37 \\ \hline
\end{tabular}
\end{center}
\end{table}

\begin{table}[htb]
\renewcommand{\tabcolsep}{3mm}
\begin{minipage}{0.4\linewidth}
\caption{\label{tab:strongdeltas}Strong phase shifts}
\centering
\begin{tabular}{|c|c|} \hline
$\delta_1$ & -0.47\\ \hline
$\delta_0$ & -0.044\\ \hline
$\delta_2$ & 0.16\\ \hline
\end{tabular}
\end{minipage}
\begin{minipage}{0.4\linewidth}
\centering
\caption{\label{tab:csbdeltas}CSB phase shifts}
\begin{tabular}{|c|c|} \hline
$\overline{\delta}_1$ & -0.44\\ \hline
$\overline{\delta}_\alpha$ & 0.19\\ \hline
$\overline{\delta}_\beta$ & -0.43\\ \hline
$\overline{\delta}_2$ & 0.44\\ \hline
\end{tabular}
\end{minipage}
\end{table}

\bibliography{references}

\begin{thebibliography}{28}
\expandafter\ifx\csname natexlab\endcsname\relax\def\natexlab#1{#1}\fi
\expandafter\ifx\csname bibnamefont\endcsname\relax
  \def\bibnamefont#1{#1}\fi
\expandafter\ifx\csname bibfnamefont\endcsname\relax
  \def\bibfnamefont#1{#1}\fi
\expandafter\ifx\csname citenamefont\endcsname\relax
  \def\citenamefont#1{#1}\fi
\expandafter\ifx\csname url\endcsname\relax
  \def\url#1{\texttt{#1}}\fi
\expandafter\ifx\csname urlprefix\endcsname\relax\def\urlprefix{URL }\fi
\providecommand{\bibinfo}[2]{#2}
\providecommand{\eprint}[2][]{\url{#2}}

\bibitem[{\citenamefont{Jenkins and Manohar}(1991)}]{Jenkins:1990jv}
\bibinfo{author}{\bibfnamefont{E.~E.} \bibnamefont{Jenkins}} \bibnamefont{and}
  \bibinfo{author}{\bibfnamefont{A.~V.} \bibnamefont{Manohar}},
  \bibinfo{journal}{Phys. Lett.} \textbf{\bibinfo{volume}{B255}},
  \bibinfo{pages}{558} (\bibinfo{year}{1991}).

\bibitem[{\citenamefont{Bernard et~al.}(1993)\citenamefont{Bernard, Kaiser, and
  Meissner}}]{Bernard:1993nj}
\bibinfo{author}{\bibfnamefont{V.}~\bibnamefont{Bernard}},
  \bibinfo{author}{\bibfnamefont{N.}~\bibnamefont{Kaiser}}, \bibnamefont{and}
  \bibinfo{author}{\bibfnamefont{U.~G.} \bibnamefont{Meissner}},
  \bibinfo{journal}{Z. Phys.} \textbf{\bibinfo{volume}{C60}},
  \bibinfo{pages}{111} (\bibinfo{year}{1993}).

\bibitem[{\citenamefont{Hemmert et~al.}(1998)\citenamefont{Hemmert, Holstein,
  and Kambor}}]{Hemmert:1997ye}
\bibinfo{author}{\bibfnamefont{T.~R.} \bibnamefont{Hemmert}},
  \bibinfo{author}{\bibfnamefont{B.~R.} \bibnamefont{Holstein}},
  \bibnamefont{and} \bibinfo{author}{\bibfnamefont{J.}~\bibnamefont{Kambor}},
  \bibinfo{journal}{J. Phys.} \textbf{\bibinfo{volume}{G24}},
  \bibinfo{pages}{1831} (\bibinfo{year}{1998}).

\bibitem[{\citenamefont{Bedaque and van Kolck}(2002)}]{Bedaque:2002mn}
\bibinfo{author}{\bibfnamefont{P.~F.} \bibnamefont{Bedaque}} \bibnamefont{and}
  \bibinfo{author}{\bibfnamefont{U.}~\bibnamefont{van Kolck}},
  \bibinfo{journal}{Ann. Rev. Nucl. Part. Sci.} \textbf{\bibinfo{volume}{52}},
  \bibinfo{pages}{339} (\bibinfo{year}{2002}).

\bibitem[{\citenamefont{Cheung et~al.}(1980)\citenamefont{Cheung, Henley, and
  Miller}}]{Cheung:1979ma}
\bibinfo{author}{\bibfnamefont{C.~Y.} \bibnamefont{Cheung}},
  \bibinfo{author}{\bibfnamefont{E.~M.} \bibnamefont{Henley}},
  \bibnamefont{and} \bibinfo{author}{\bibfnamefont{G.~A.}
  \bibnamefont{Miller}}, \bibinfo{journal}{Nucl. Phys.}
  \textbf{\bibinfo{volume}{A348}}, \bibinfo{pages}{365} (\bibinfo{year}{1980}).

\bibitem[{\citenamefont{Epelbaum and Meissner}(1999)}]{Epelbaum:1999zn}
\bibinfo{author}{\bibfnamefont{E.}~\bibnamefont{Epelbaum}} \bibnamefont{and}
  \bibinfo{author}{\bibfnamefont{U.-G.} \bibnamefont{Meissner}},
  \bibinfo{journal}{Phys. Lett.} \textbf{\bibinfo{volume}{B461}},
  \bibinfo{pages}{287} (\bibinfo{year}{1999}).

\bibitem[{\citenamefont{Miller et~al.}(2006)\citenamefont{Miller, Opper, and
  Stephenson}}]{Miller:2006tv}
\bibinfo{author}{\bibfnamefont{G.~A.} \bibnamefont{Miller}},
  \bibinfo{author}{\bibfnamefont{A.~K.} \bibnamefont{Opper}}, \bibnamefont{and}
  \bibinfo{author}{\bibfnamefont{E.~J.} \bibnamefont{Stephenson}},
  \bibinfo{journal}{Ann. Rev. Nucl. Part. Sci.} \textbf{\bibinfo{volume}{56}},
  \bibinfo{pages}{253} (\bibinfo{year}{2006}).

\bibitem[{\citenamefont{Slaus et~al.}(1991)\citenamefont{Slaus, Nefkens, and
  Miller}}]{Slaus:1990nn}
\bibinfo{author}{\bibfnamefont{I.}~\bibnamefont{Slaus}},
  \bibinfo{author}{\bibfnamefont{B.~M.~K.} \bibnamefont{Nefkens}},
  \bibnamefont{and} \bibinfo{author}{\bibfnamefont{G.~A.}
  \bibnamefont{Miller}}, \bibinfo{journal}{Nucl. Instrum. Meth.}
  \textbf{\bibinfo{volume}{B56-57}}, \bibinfo{pages}{489}
  (\bibinfo{year}{1991}).

\bibitem[{\citenamefont{Weinberg}(1977)}]{Weinberg:1977}
\bibinfo{author}{\bibfnamefont{S.}~\bibnamefont{Weinberg}},
  \bibinfo{journal}{Trans. N.Y. Acad. Sci.} \textbf{\bibinfo{volume}{38}},
  \bibinfo{pages}{185} (\bibinfo{year}{1977}).

\bibitem[{\citenamefont{Opper et~al.}(2003)}]{Opper:2003sb}
\bibinfo{author}{\bibfnamefont{A.~K.} \bibnamefont{Opper}}
  \bibnamefont{et~al.}, \bibinfo{journal}{Phys. Rev. Lett.}
  \textbf{\bibinfo{volume}{91}}, \bibinfo{pages}{212302}
  (\bibinfo{year}{2003}).

\bibitem[{\citenamefont{Lensky et~al.}(2006)}]{Lensky:2005jc}
\bibinfo{author}{\bibfnamefont{V.}~\bibnamefont{Lensky}} \bibnamefont{et~al.},
  \bibinfo{journal}{Eur. Phys. J.} \textbf{\bibinfo{volume}{A27}},
  \bibinfo{pages}{37} (\bibinfo{year}{2006}).

\bibitem[{\citenamefont{Niskanen}(1999)}]{Niskanen:1998yi}
\bibinfo{author}{\bibfnamefont{J.~A.} \bibnamefont{Niskanen}},
  \bibinfo{journal}{Few Body Syst.} \textbf{\bibinfo{volume}{26}},
  \bibinfo{pages}{241} (\bibinfo{year}{1999}).

\bibitem[{\citenamefont{van Kolck et~al.}(2000)\citenamefont{van Kolck,
  Niskanen, and Miller}}]{vanKolck:2000ip}
\bibinfo{author}{\bibfnamefont{U.}~\bibnamefont{van Kolck}},
  \bibinfo{author}{\bibfnamefont{J.~A.} \bibnamefont{Niskanen}},
  \bibnamefont{and} \bibinfo{author}{\bibfnamefont{G.~A.}
  \bibnamefont{Miller}}, \bibinfo{journal}{Phys. Lett.}
  \textbf{\bibinfo{volume}{B493}}, \bibinfo{pages}{65} (\bibinfo{year}{2000}).

\bibitem[{\citenamefont{Filin et~al.}(2009)}]{Filin:2009yh}
\bibinfo{author}{\bibfnamefont{A.}~\bibnamefont{Filin}} \bibnamefont{et~al.}
  (\bibinfo{year}{2009}), \eprint{0907.4671}.

\bibitem[{\citenamefont{Weinberg}(1991)}]{Weinberg:1991um}
\bibinfo{author}{\bibfnamefont{S.}~\bibnamefont{Weinberg}},
  \bibinfo{journal}{Nucl. Phys.} \textbf{\bibinfo{volume}{B363}},
  \bibinfo{pages}{3} (\bibinfo{year}{1991}).

\bibitem[{\citenamefont{Wiringa et~al.}(1995)\citenamefont{Wiringa, Stoks, and
  Schiavilla}}]{Wiringa:1994wb}
\bibinfo{author}{\bibfnamefont{R.~B.} \bibnamefont{Wiringa}},
  \bibinfo{author}{\bibfnamefont{V.~G.~J.} \bibnamefont{Stoks}},
  \bibnamefont{and}
  \bibinfo{author}{\bibfnamefont{R.}~\bibnamefont{Schiavilla}},
  \bibinfo{journal}{Phys. Rev.} \textbf{\bibinfo{volume}{C51}},
  \bibinfo{pages}{38} (\bibinfo{year}{1995}).

\bibitem[{\citenamefont{Hanhart}(2004)}]{Hanhart:2003pg}
\bibinfo{author}{\bibfnamefont{C.}~\bibnamefont{Hanhart}},
  \bibinfo{journal}{Phys. Rept.} \textbf{\bibinfo{volume}{397}},
  \bibinfo{pages}{155} (\bibinfo{year}{2004}).

\bibitem[{\citenamefont{Hanhart et~al.}(1995)\citenamefont{Hanhart,
  Haidenbauer, Reuber, Schutz, and Speth}}]{Hanhart:1995ut}
\bibinfo{author}{\bibfnamefont{C.}~\bibnamefont{Hanhart}},
  \bibinfo{author}{\bibfnamefont{J.}~\bibnamefont{Haidenbauer}},
  \bibinfo{author}{\bibfnamefont{A.}~\bibnamefont{Reuber}},
  \bibinfo{author}{\bibfnamefont{C.}~\bibnamefont{Schutz}}, \bibnamefont{and}
  \bibinfo{author}{\bibfnamefont{J.}~\bibnamefont{Speth}},
  \bibinfo{journal}{Phys. Lett.} \textbf{\bibinfo{volume}{B358}},
  \bibinfo{pages}{21} (\bibinfo{year}{1995}).

\bibitem[{\citenamefont{Hutcheon et~al.}(1990)}]{Hutcheon:1989bt}
\bibinfo{author}{\bibfnamefont{D.~A.} \bibnamefont{Hutcheon}}
  \bibnamefont{et~al.}, \bibinfo{journal}{Phys. Rev. Lett.}
  \textbf{\bibinfo{volume}{64}}, \bibinfo{pages}{176} (\bibinfo{year}{1990}).

\bibitem[{\citenamefont{Gardestig et~al.}(2006)\citenamefont{Gardestig,
  Phillips, and Elster}}]{Gardestig:2005sn}
\bibinfo{author}{\bibfnamefont{A.}~\bibnamefont{Gardestig}},
  \bibinfo{author}{\bibfnamefont{D.~R.} \bibnamefont{Phillips}},
  \bibnamefont{and} \bibinfo{author}{\bibfnamefont{C.}~\bibnamefont{Elster}},
  \bibinfo{journal}{Phys. Rev.} \textbf{\bibinfo{volume}{C73}},
  \bibinfo{pages}{024002} (\bibinfo{year}{2006}).

\bibitem[{\citenamefont{Baru et~al.}(2009)}]{Baru:2009fm}
\bibinfo{author}{\bibfnamefont{V.}~\bibnamefont{Baru}} \bibnamefont{et~al.}
  (\bibinfo{year}{2009}), \eprint{0907.3911}.

\bibitem[{\citenamefont{Hanhart et~al.}(2000)\citenamefont{Hanhart, van Kolck,
  and Miller}}]{Hanhart:2000gp}
\bibinfo{author}{\bibfnamefont{C.}~\bibnamefont{Hanhart}},
  \bibinfo{author}{\bibfnamefont{U.}~\bibnamefont{van Kolck}},
  \bibnamefont{and} \bibinfo{author}{\bibfnamefont{G.~A.}
  \bibnamefont{Miller}}, \bibinfo{journal}{Phys. Rev. Lett.}
  \textbf{\bibinfo{volume}{85}}, \bibinfo{pages}{2905} (\bibinfo{year}{2000}).

\bibitem[{\citenamefont{Heimberg et~al.}(1996)}]{Heimberg:1996be}
\bibinfo{author}{\bibfnamefont{P.}~\bibnamefont{Heimberg}}
  \bibnamefont{et~al.}, \bibinfo{journal}{Phys. Rev. Lett.}
  \textbf{\bibinfo{volume}{77}}, \bibinfo{pages}{1012} (\bibinfo{year}{1996}).

\bibitem[{\citenamefont{Weinberg}(1994)}]{Weinberg:1994tu}
\bibinfo{author}{\bibfnamefont{S.}~\bibnamefont{Weinberg}}
  (\bibinfo{year}{1994}), \eprint{hep-ph/9412326}.

\bibitem[{\citenamefont{Gasser and Leutwyler}(1982)}]{Gasser:1982ap}
\bibinfo{author}{\bibfnamefont{J.}~\bibnamefont{Gasser}} \bibnamefont{and}
  \bibinfo{author}{\bibfnamefont{H.}~\bibnamefont{Leutwyler}},
  \bibinfo{journal}{Phys. Rept.} \textbf{\bibinfo{volume}{87}},
  \bibinfo{pages}{77} (\bibinfo{year}{1982}).

\bibitem[{\citenamefont{van Kolck et~al.}(1996)\citenamefont{van Kolck, Friar,
  and Goldman}}]{vanKolck:1996rm}
\bibinfo{author}{\bibfnamefont{U.}~\bibnamefont{van Kolck}},
  \bibinfo{author}{\bibfnamefont{J.~L.} \bibnamefont{Friar}}, \bibnamefont{and}
  \bibinfo{author}{\bibfnamefont{J.~T.} \bibnamefont{Goldman}},
  \bibinfo{journal}{Phys. Lett.} \textbf{\bibinfo{volume}{B371}},
  \bibinfo{pages}{169} (\bibinfo{year}{1996}).

\bibitem[{\citenamefont{Coon and Scadron}(1995)}]{Coon:1995xp}
\bibinfo{author}{\bibfnamefont{S.~A.} \bibnamefont{Coon}} \bibnamefont{and}
  \bibinfo{author}{\bibfnamefont{M.~D.} \bibnamefont{Scadron}},
  \bibinfo{journal}{Phys. Rev.} \textbf{\bibinfo{volume}{C51}},
  \bibinfo{pages}{2923} (\bibinfo{year}{1995}).

\bibitem[{\citenamefont{Cohen et~al.}(1996)\citenamefont{Cohen, Friar, Miller,
  and van Kolck}}]{Cohen:1995cc}
\bibinfo{author}{\bibfnamefont{T.~D.} \bibnamefont{Cohen}},
  \bibinfo{author}{\bibfnamefont{J.~L.} \bibnamefont{Friar}},
  \bibinfo{author}{\bibfnamefont{G.~A.} \bibnamefont{Miller}},
  \bibnamefont{and} \bibinfo{author}{\bibfnamefont{U.}~\bibnamefont{van
  Kolck}}, \bibinfo{journal}{Phys. Rev.} \textbf{\bibinfo{volume}{C53}},
  \bibinfo{pages}{2661} (\bibinfo{year}{1996}).

\end{thebibliography}

\end{document}